\documentclass[%
 reprint,
superscriptaddress,
 amsmath,amssymb,
 aps,
floatfix,
]{revtex4-2}

\usepackage{graphicx}
\usepackage{dcolumn}
\usepackage{physics}
\usepackage{bm}

\usepackage[caption = false]{subfig}
\usepackage{float}
\begin{document}

\preprint{APS/123-QED}

\title{Effect of Electric Fields on the Director Field and Shape of Nematic Tactoids }

\author{Mohammadamin Safdari}
\author{Roya Zandi}%
\affiliation{%
Department of Physics, University of California, Riverside, California 92521, USA
}%
\author{Paul van der Schoot}
\affiliation{
Department of Applied Physics, Eindhoven University of Technology, The Netherlands\\
}%


\date{\today}

\begin{abstract}
Tactoids are spindle-shaped droplets of a uniaxial nematic phase suspended in the co-existing isotropic phase. They are found in dispersions of a wide variety of elongated colloidal particles, including actin, fd virus, carbon nanotubes, vanadium peroxide and chitin nanocrystals. Recent experiments on tactoids of chitin nanocrystals in water show that electric fields can very strongly elongate tactoids even though the dielectric properties of the co-existing isotropic and nematic phases differ only subtly. We develop a model for
partially bipolar tactoids, where the degree of bipolarness of the director field is free to adjust to optimize the sum of the elastic, surface and Coulomb energies of the system. By means of a combination of a scaling analysis and a numerical study, we investigate the elongation and director field's behavior of the tactoids as a function of their size, the strength of the electric field, the surface tension, anchoring strength, the elastic constants and the electric susceptibility anisotropy. We find that tactoids cannot elongate significantly due to an external electric field, unless the director field is bipolar or quasi bipolar and is somehow frozen in the field-free configuration. Presuming that this is the case, we find reasonable agreement with experimental data.

\end{abstract}

\maketitle


\section{\label{sec:level1}Introduction 
}
If an isotropic phase of rod-like colloidal particles undergoes a phase transition leading to a coexisting, uniaxially ordered nematic phase, then this typically happens via an intermediate stage characterized by an isotropic background phase in which are dispersed spindle-shaped droplets called tactoids, see Fig 1. These tactoids eventually sediment and coalesce to form a macroscopic nematic phase, although this may take a very long time \cite{van1992,van2005,Lettinga2005,Lettinga2006,Jamali2015,fraden1985}. First discovered in 1925 by Zocher, who also coined the term tactoids (``Taktoide'' in German), in vanadium pentoxide sols  \cite{Zocher1925}, they have since been observed in a plethora of molecular, polymeric and colloidal lyotropic liquid crystals. These include dispersions of tobacco mosaic virus particles \citep{BAWDEN1936}, iron oxyhydroxide nanorods \cite{Coper1937}, polypeptides \cite{Robinson1956}, carbon nanotubes \cite{puech2010,Jamali2015,jamali2017}, fd virus particles \cite{dogic2003,modlinska2015}, F-actin fibers \cite{oakes2007}, actin filaments\cite{weirich2017liquid,weirich2019self} , chromonic liquid crystals \cite{Kim2013}, amyloid fibers \cite{marenduzzo2005} and cellulose nanocrystals \cite{kitzerow1994,park2014,wang2016,REVOL1992}. 

The peculiar, pointy and elongated shape of the tactoids, which reflects the underlying symmetry of the nematic phase, was initially explained in terms of the surface anchoring of the director field, presumed to be uniform \cite{chandrasekhar1966}. The fact that the degree of elongation depends on the volume of the droplets, and that polarization microscopic images show them to be bipolar rather than uniform, at least if they are sufficiently large \cite{Jamali2015}, reveals that this explanation is incomplete. In a bipolar configuration the director field conforms to a bi-spherical coordinate system, illustrated in Fig. 1 and in more detail in Fig. 2. If the focal points of the coordinate system reside on the poles of the droplets, representing proper surface defects known as boojums \cite{lavrentovich1998}, the director field is then  properly bipolar.

Theoretical studies of Kaznacheev \textsl{et al.}  \cite{Kaznacheev2002,Kaznacheev2003} and Prinsen \textsl{et al.} \cite{Prinsen2003,Prinsen2004a,Prinsen2004b} have revealed that the boojums are by and large virtual, situated outside of the droplet in an extrapolated director field pattern, and that the director field is almost always incompletely or quasi bipolar  \cite{Kaznacheev2002,Prinsen2003,Lettinga2006,van2012,otten2012,Everts2016}. The smaller the droplet, the further the virtual boojums move away from the poles of the droplets and the more strongly the director field resembles that of a spatially uniform director field that represents its ground state. The full crossover from uniform to bipolar director fields has only recently been observed experimentally for tactoids in dispersions of carbon nanotubes in chlorosulfonic acid, both in bulk and sessile, that is, on planar surfaces \cite{Jamali2015,jamali2017}, see Fig. 1.

What has emerged, is a picture in which there are two length scales that predict the structure and shape of tactoids of a certain size. Following de Gennes, these length scales may perhaps be called extrapolation lengths, and are defined as ratios of elastic constants and surface energies \cite{de1993}. These surface energies are the bare surface tension between the isotropic and nematic phases, and the surface anchoring energy penalising a deviation from the preferred planar anchoring of the director field of elongated colloidal particles along the interface \cite{van1999}. Droplets that are smaller than the smallest of these two length scales, that is, the length scale associated with the surface anchoring, tend to have a uniform director field and elongated shape. If a droplet is larger than the larger extrapolation length, which is associated with the bare surface tension, then it tends to be bipolar and nearly spherical. Droplets of a size in between these two length scales remain elongated but have director field in between uniform and bipolar, see Fig.~1.

The situation becomes more complex if yet another length scale enters the stage. For instance, if the nematic is not uniaxial but chiral, that is, cholesteric, then the cholesteric pitch interferes with these two length scales. This gives rise to an additional regime separating uniaxial from twisted nematic (cholesteric) configurations \cite{marenduzzo2004,Prinsen2004b,Williams1986,vanzo2012shape}. The same is true if an external electric or magnetic field is applied to nematic rather than cholesteric tactoids. The impact of a magnetic field on tactoids of vanadium pentoxide fibers dispersed in water was investigated experimentally and theoretically by Kaznacheev \textit{et al.}, who found that an externally applied magnetic field stretches tactoids, at least if they are sufficiently large \cite{Kaznacheev2002}. This, indeed, points at the existence of another pertinent length scale.

The existence of such a length scale was recently confirmed  by Metselaar \textit{et al.}, who studied the impact of a high-frequency electric AC field on tactoids of chitin fibers dispersed in water \cite{Metselaar2017}. These authors find very large elongations of tactoids in the presence of an electric field, with aspect ratios increasing from about two in zero field to about twenty for droplets larger than some critical size. Their numerical simulations, based on the lattice Boltzmann method, mimic this observation, showing that in order to obtain a very large length-to-breath or aspect ratio for the droplets, a large anchoring strength is required. Interestingly, large anchoring strengths are also known to lead to quite elongated tactoids in zero field, but the effect is apparently somehow dramatically enhanced by an electric field that arguably align the fibers and hence also the director field along the field direction. 

If the planar anchoring of the director field to the interface between the coexisting isotropic and nematic phases were absolutely rigid and strictly bipolar, then the theory of Kaznacheev and collaborators \cite{Kaznacheev2002} would predict an in principle unbounded growth of the length of the tactoids with increasing electric or magnetic field strength. \footnote{Tactoids can also be stretched by the effect an elongational flow field, as was recently shown by H. Almohammadi, M. Bagnani and R. Mezzenga, Flow-induced order–order transitions in amyloid fibril liquid crystalline tactoids, Nat. Comm. 11 (2020), 5416.}

Actually, the chitin tactoids are not actually strictly bipolar but quasi bipolar, in which case the anchoring would be imperfect. 

Interestingly, in the lattice Boltzmann simulations of Metselaar {\it et al.} the director field seems to respond to the external alignment field not by keeping the bi-spherical geometry and simply stretching it, as is presumed in the calculations of Kaznacheev and co-workers. Instead, the director field seems to become uniform in the center of the droplet to bend sharply close to the interface in order to accommodate planar anchoring \cite{Metselaar_private}. This is highly surprising, because such a change in the geometry of the director field would require very large local deformation of the director field and associated with that would be a large elastic free energy of deformation. We should perhaps not exclude the possibility that the limited spatial resolution of the simulations produces such a strong director field deformation \cite{Metselaar_private}.

In this paper we delve more deeply into the problem of how external fields deform nematic tactoids, extending the theory of Kaznacheev \textsl{et al.} \cite{Kaznacheev2002} by allowing for imperfect anchoring. By means of a combination of a scaling analysis and a numerical minimization of a free energy with prescribed director field geometry and droplet shape, we obtain that the external field cannot produce aspect ratios that exceed those in zero field. We find a highly complex behavior characterized by no fewer than five different scaling regimes for the elongation and director field of tactoids. If we fix the geometry of the equilibrium director field at zero field, and let only the aspect ratio respond to the external field, we do find very large aspect ratios for large field strengths. 

Our predictions agree qualitatively with the experimental findings of Metselaar {\it et al.} \cite{Metselaar2017}. This seems to suggest that the droplet shape and the director field relax with different rates in response to an external field that is suddenly switched on. In follow-up work, we intend to study a two-mode relaxational dynamics model to investigate further the observations of Metselaar \textsl{et al.} \cite{Metselaar2017}.

\begin{figure}[!ht]
	\begin{center}
		\includegraphics[width=8cm]{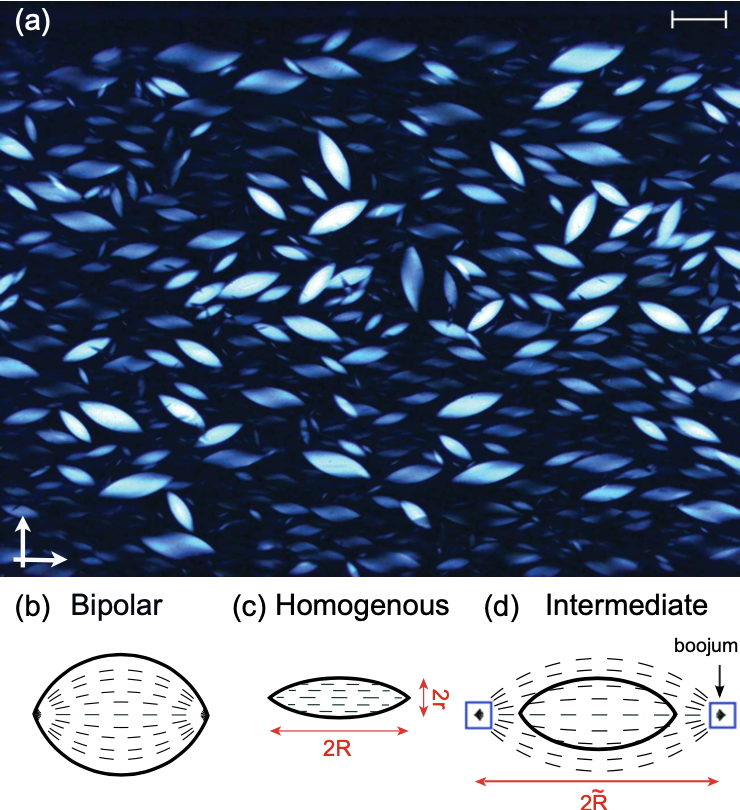}
		\caption{(a) Polarized optical micrograph illustrating a number of nematic tactoids of different size and extinction pattern associated with the director-field conformation, in a solution of carbon nanotubes in chlorosulfonic acid at 1000 ppm. Arrows show the orien- tation of crossed polarizers. Schematics of (b) a bipolar tactoid with boojum surface defects at the poles, (c) a homogenous tactoid, and (d) an intermediate tactoid described by virtual boojums outside of the droplet. The parameters $R$ and $r$ represent the major and minor axes of the tactoid, respectively. Adopted from Ref. \cite{Jamali2015}.} 
		 \centering 
		\label{fig1}
	\end{center}
\end{figure}

The remainder of the paper is structured as follows. In Section II, we present our free energy that consists of a contribution of the Oseen-Frank elastic free energy, a Rapini-Papoular surface free energy and a Coulomb free energy associated with the electric field. In Section III, we work out the scaling theory of fully bipolar and quasi-bipolar droplets in the presence of an external field, producing the various relevant length scales in the problem. 
 
In section IV, we compare the results of the variational theory that we evaluate numerically with our findings of the scaling theory, and find a very good agreement. Finally, in Section V we summarize our findings, compare our predictions with the experimental results of Metselaar \textsl{et al.} \cite{Metselaar2017} and discuss the potential implications for our understanding of the relaxation dynamics of nematic tactoids.

\section{Free Energy}
We consider a nematic droplet suspended in an isotropic fluid medium. The free energy $F$ describing the droplet in an external electrical field consists of a sum of three terms, 
\begin{equation}
F = F_\mathrm{E} + F_\mathrm{S} + F_\mathrm{C},
\label{TotalFreeEnergy}
\end{equation}
representing the Frank elastic free energy associated with a potentially deformed director field $F_\mathrm{E}$, an interfacial free energy $F_\mathrm{S}$, and a Coulomb energy $F_\mathrm{C}$.

Focusing on twist-free bipolar director fields, the Frank elastic free energy of the droplet reads \cite{schiller1996}, 
\begin{equation}
\begin{aligned}
F_\mathrm{E} =  &\int  \left[  \dfrac{1}{2} K_{11} (\vec{\nabla} \cdot \vec{n})^2 + \dfrac{1}{2} K_{33} \left(\vec{n} \times (\vec{\nabla} \times \vec{n})\right)^2 \right.
\\
& \quad \left. - \dfrac{1}{2} K_{24} \vec{\nabla} \cdot \left( \vec{n} \vec{\nabla} \cdot \vec{n} + \vec{n} \times (\vec{\nabla} \times \vec{n}) \right) \right] \, \mathrm{d} V ,
\end{aligned}
\label{FrankFreeEnergy}
\end{equation}
where the integration is over the entire volume $V$ of the droplet, $\vec{n}$ represents the position-dependent director field, and $K_{11}$, $K_{33}$ and $K_{24}$ are the elastic moduli of the splay, bend and saddle-splay deformations, respectively \cite{de1993}. Here, we do not allow for twisted director fields that may arise if the bend elastic constant is sufficiently small \citep{Williams1986}. Note that these parity-broken structures are anyway suppressed if the tactoids are elongated \cite{Prinsen2004b}.

Within a Rapini-Papoular approximation \cite{rapini1969}, the interfacial free energy can be written as 
\begin{equation}
F_\mathrm{S} = \sigma \int \left[(1+\omega \left(\vec{q} \cdot \vec{n}\right)^2 \right] \, \mathrm{d}A,
\label{InterfaceFreeEnergy}
\end{equation}
where $\sigma$ is the interfacial tension between the nematic phase of the droplet and isotropic medium, $\omega$ is a dimensionless anchoring strength and the integration is over the interfacial area $A$ of the droplet. We presume that $\omega>0$, implying that the anchoring penalises a director field $\vec{n}$ that is not parallel to the interface, that is, at right angles to the surface normal $\vec{q}$. Rod-like particles prefer planar anchoring of the nematic at the interface with the coexisting isotropic phase for entropy reasons \cite{van2005,chen1992,van1999}. In principle, both the surface tension and anchoring strength could depend on the curvature of the interface, but even for very small droplets the effect seems to be very small \cite{Everts2016}.

Finally, the Coulomb energy of a nematic droplet in an electric field $\vec{E}$ can be written as \cite{de1993,landau2013}, 
\begin{equation}
F_\mathrm{C}= - \dfrac{1}{8 \pi} \epsilon_{a} \int \left(\vec{n} \cdot \vec{E} \right)^2 \, \mathrm{d}V,
\label{ElectricFieldFreeEnergy}
\end{equation}
where $\epsilon_{a}=\epsilon_{\parallel}-\epsilon_{\perp} \geq 0 $ is the dielectric susceptibility anisotropy of the dispersion of rod-like particles, which can be described as a second-rank tensor with components $\epsilon_{\parallel}$ and $\epsilon_{\perp}$ parallel and perpendicular to the droplet axis \cite{doi1988}. Note that we ignore a potential permanent dipole moment on the particles and that we have not explicitly written a constant term that is not a function of the director field. It is important to note that both $\epsilon_{\parallel}$ and $\epsilon_{\perp}$ are not all that different from the dielectric constant of the isotropic phase, because the dielectric response of the suspension is dominated by that of the solvent \cite{Metselaar2017}. This means that any elongation of the droplets caused by an electric field is not due to a difference between the dielectric properties of the isotropic and nematic phases, as would be the case for a thermotropic nematic tactoid suspended in a polymeric fluid \cite{lev2000}, but due to the anisotropy of the dielectric response of the nematic phase itself.

Having collected all contributions to the free energy, we need to address an issue of some contention, which is whether or not the susceptibility anisotropy, the surface energies and the elastic constants depend on the strength of the electric field. In principle, they do. The reason is that these quantities depend on the level of alignment of the particles in the coexisting isotropic and nematic phases \cite{straley1973,odijk1986,van1999,vroege1992}. We also note that strictly speaking the isotropic phase becomes paranematic in the presence of an alignment field. In fact, the isotropic-to-nematic phase transition ends in a critical point, at which both the interfacial tension and anchoring should vanish \cite{fraden1985,wensink2003,khokhlov1982,semenov1999}. To keep our analysis as simple as possible, and within the philosophy of linear response theory, we shall ignore any impact of the electric field on the elastic constants and surface energies, and presume the external field in some sense to be sufficiently weak not to affect these quantities yet sufficiently strong to deform the tactoids. 

To find the equilibrium shape and director-field configuration of tactoids, we would need to solve the appropriate Euler-Lagrange equations that result from a minimization of the free energy, given in Eq.~\ref{TotalFreeEnergy}. The minimization is with respect to the director field $\vec{n}\left(\vec{r}\right)$, which depends on the spatial coordinate $\vec{r}$, as well as on the droplet shape. This has to be done subject to the conditions of a constant droplet volume and a constant unit length of the director,  $|\vec{n}|\equiv 1$ \cite{schiller1996}, which produces a quite complex mathematical problem, also numerically, in view of the free boundary \cite{Williams1986,Prinsen2003}.

Hence, we follow the earlier work of Kaznacheev \textsl{et al.} \cite{Kaznacheev2002,Kaznacheev2003} and Prinsen \textsl{et al.} \cite{Prinsen2003,Prinsen2004a,Prinsen2004b}, and restrict geometry of both the director field and droplet shape. For the shapes of the droplets we use circle sections rotated about their chord, producing potentially elongated droplets with sharp ends that are very similar to the tactoid shapes found in a wide variety of experiments, including those of Metselaar \textsl{et al.} \cite{Metselaar2017}. We are aware that the equilibrium shape of the poles is a cusp if $0 \leq \omega <1$ and the director field (nearly) uniform, which in that case has a more rounded form. However, as we showed in \cite{Prinsen2003}, the free energy difference between the exact Wulff shape, the spindle shape and ellipsoids of revolution is minute for $\omega <1$, so we deem the approximation not to be a grave one.

\begin{figure}
    \centering
    \includegraphics[width=8cm]{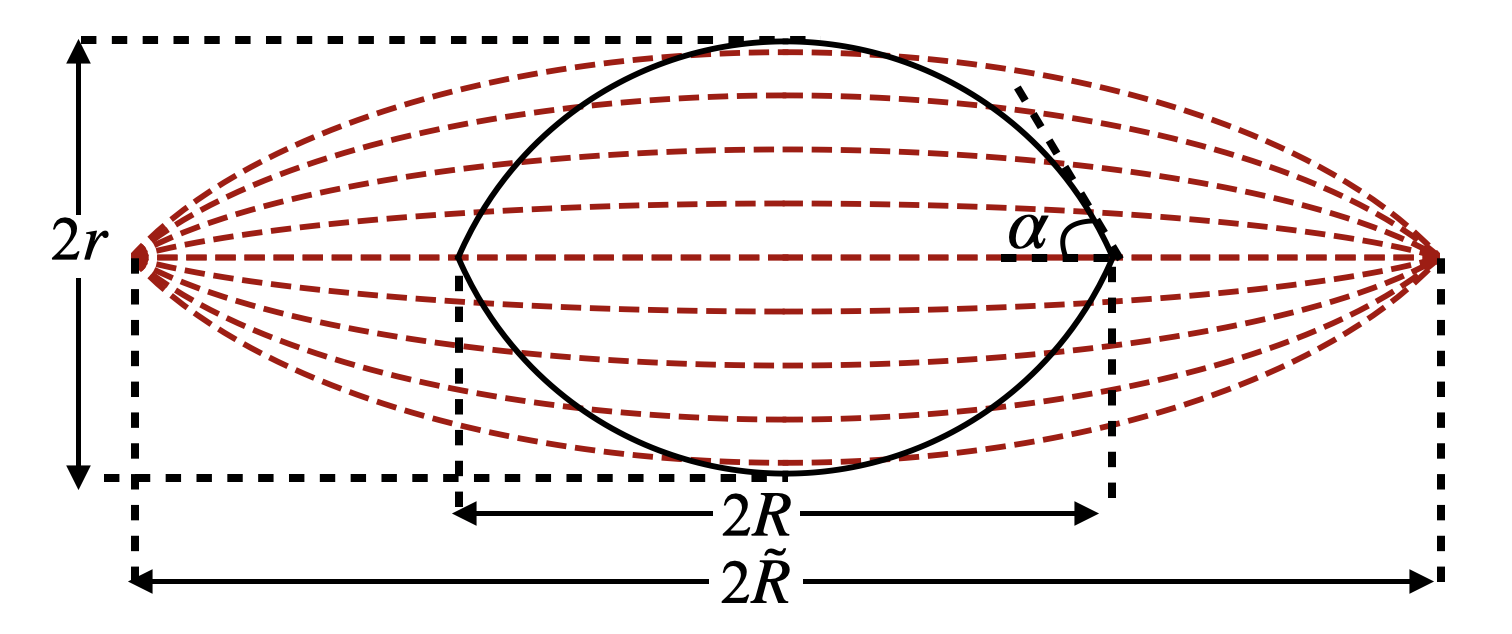}
    \caption{Cross section (solid) and director field (dashed) of a tactoid presumed in our calculations. The droplet is cylindrically symmetric about its main axis. $R$ denotes the length of a tactoid and $r$ its width. $\tilde{R}$ is the distance between the virtual boojums, which are the focal points of the (extrapolated) director field. For $R=\tilde{R}$, the virtual boojums become actual boojums, {\it i.e.}, surface point defects. Also indicated is $\alpha$, the opening angle of the spindle-shaped droplet, see also the main text.
    }
    \label{fig2-1}
\end{figure}

For the director field we employ a bispherical coordinate system first used by Williams to describe spherical bipolar droplets \cite{Williams1986}, and more recently by Prinsen \textsl{et al.} \cite{Prinsen2003,Prinsen2004a,Prinsen2004b}, Jamali \textsl{et al.} \cite{Jamali2015,jamali2017} and  Kaznacheev \textsl{et al.} \cite{Kaznacheev2002,Kaznacheev2003} for elongated bipolar droplets. We do not fix the position of the foci of the bispherical director field to the poles of the tactoid to allow for a smooth interpolation between a uniform and bipolar director field, although we do prescribe them to reside on the main axis of revolution of the tactoid. See also Fig.~\ref{fig2-1}. 

Within this prescription, the shape and director-field configuration of a tactoid is completely described by two parameters, at least if the volume of the droplet is known. These two parameters are the opening angle $\alpha$ and the ratio $y\equiv \tilde{R}/R$ of the distance between the virtual boojums $2\tilde{R}$ and length of the droplet $2R$. The former quantity describes its aspect ratio $x \equiv R/r = \cot (\alpha/2)$, with $r$ the width of the droplet, and the latter the degree of ``bipolarness'' of the director field. For a spherical droplet, we have $x = 1$ and $\alpha = \pi/2$, whilst for a strongly elongated one $x \gg 1$ and $\alpha \ll 1$. For a tactoid with a uniform director field $y \gg 1$, and for a bipolar tactoid $y\rightarrow 1$, see also Fig.~\ref{fig2-1}. We note that the quantities $x$ and $y$ will explicitly contribute to the scaling theory presented in the following section.

It turns out practical to render the free energy dimensionless, and define $f = f(\alpha, y) \equiv \sigma F/\left(K_{11}-K_{24}\right)^2$. The optimal free energy minimizes $f$ with respect to the opening angle $\alpha$ or aspect ratio $x$ and the bipolarness $y$, keeping the volume $V$ of the droplet constant. Let the dimensionless volume be defined as $v\equiv V\left(\sigma/(K_{11}-K_{24})\right)^3$. The dimensionless free energy $f$ can then be written as a sum of surface and volume terms 
\begin{equation}
\begin{aligned}
 f(\alpha,y) &  =  v^{2/3} \phi_\mathrm{v}^{-2/3}(\alpha) \left[ \phi_{\sigma}(\alpha) +\omega  \phi_{\omega} (\alpha,y) \right] \\
& + v^{1/3} \phi_\mathrm{v}^{-1/3}(\alpha) \left[ \phi_{11}(\alpha,y)
 + \kappa \phi_{33}(\alpha,y) \right] \\
& -  v \phi_\mathrm{v}^{-1}(\alpha) \Gamma \phi_{\mathrm{C}}(\alpha,y),
\label{eq5}
\end{aligned}
\end{equation}
where we refer to the Supplementary Information (SI) for details. The first line represents the two surface contributions, the second line corresponds to the three types of elastic deformation of the director field, and the last is related to the Coulomb energy. Here, $\kappa \equiv K_{33}/(K_{11}-K_{24})$ measures the magnitude of the bend elastic constant relative to the effective splay constant, and $\Gamma=\frac{1}{8\pi} \sigma^{-2} \epsilon_a E^2 \left(K_{11}-K_{24}\right)$ is the appropriate quantity to probe the impact of the electric field relative to the surface tension and elastic deformation.

For the case of lyotropic nematics of rod-like particles, we typically have $\sigma \approx 10^{-7} - 10^{-6} $ N m$^{-1}$ for the surface tension \cite{chen2002}, and $K_{33}/K_{11} \approx 1 - 10^2$ 1 and $K_{11} \approx 10^{-12} - 10^{-11}$ N for the elastic constants \cite{jamali2017, generalova2001, Prinsen2003,sato1996,dietrich2020}. Dimensionless anchoring strengths $\omega$ are typically in the range from about 1 to 10 \cite{Prinsen2004a, Jamali2015, Kaznacheev2003, puech2010}.

All three terms are renormalized by a function measuring how the opening angle $\alpha$ affects the droplet volume for a given aspect ratio $x$. The common factor is given by
\begin{equation}
\phi_\mathrm{v}(\alpha) = \frac{7\pi}{3}+ \frac{\pi}{2}\left( \frac{1 - 4\alpha \cot \alpha + 3\cos 2\alpha}{\sin^2 \alpha}\right).
\end{equation} 
The first line of Eq.~\ref{eq5} consists of the sum of a contribution from the bare surface tension,
\begin{equation}
\begin{aligned}
\phi_\sigma(\alpha) = 4\pi \left(  \frac{1-\alpha \cot\alpha}{\sin \alpha}\right),
\end{aligned}
\end{equation}
and a term originating from the anchoring of the director field to the interface,
\begin{equation}
\begin{aligned}
\phi_{\omega}(\alpha,y) & = \frac{\pi}{2} (y^2-1)^2 \sin^3 \alpha \\ & \times \int_{0}^{\pi} \mathrm{d} \xi \left[  \dfrac{\sin^2 \xi  \cos^2 \xi   }{N(y,\xi,\alpha)\left(1+\sin \xi \cos \alpha \right)^3} \right],
\end{aligned}
\label{eqanchoring}
\end{equation}
for which we have not been able to obtain an explicit expression. Here,
\begin{equation}
\begin{aligned}
N(y,\xi,\eta)  =  & \left(\sin \xi  \cos \eta + \frac{1}{2} Z(\xi,\eta) \left( y^2 - 1 \right) \right)^2 \\ & + y^2 \sin^2 \xi \sin^2 \eta,
\end{aligned}
\label{eqN}
\end{equation}
and
\begin{equation}
Z(\xi,\eta) =  1 + \sin \xi \cos \eta.
\end{equation}
Note that in Eq.~\ref{eqanchoring}, we inserted $\eta = \alpha$ to obtain the expression for $N$ in Eq.~\ref{eqN}.

The contribution of the splay and saddle-splay deformation to the Frank elastic energy also gives rise to an integral that we have also not been able to solve analytically,
\begin{equation}
\begin{aligned}
\phi_{11}(\alpha,y) =& 8\pi  \int_{0}^{\pi} \mathrm{d} \xi \int_{0}^{\alpha} \mathrm{d} \eta \sin^2 \xi  \cos^2 \xi  \sin \eta \\
&\times \frac{1}{N(y,\xi,\eta)\left(1+\sin \xi \cos \eta \right)^3},
\end{aligned}
\end{equation}
where we note that within our family of bispherical director fields, the saddle-splay deformation merely renormalizes the contribution from the splay deformation, giving rise to an effective splay constant that is the difference of the splay and bend-splay constants, explaining the scaling of the free energy that we introduced above in terms of this difference \cite{Prinsen2003,Prinsen2004a}. The contribution of the bend elastic deformation reads
\begin{equation}
\begin{aligned}
\phi_{33}(\alpha,y) =& 8\pi \int_{0}^{\pi} \mathrm{d} \xi \int_{0}^{\alpha} \mathrm{d} \eta   \sin^4 \xi  \sin^3 \eta \\ & \times \frac{1}{N(y,\xi,\eta)\left(1+\sin \xi \cos \eta \right)^3}.
\end{aligned}
\end{equation}

Finally, the free energy of the interaction of the nematic droplet with an electric field yields an even more daunting integral,
\begin{equation}
\begin{aligned}
 \phi_{\mathrm{C}}(\alpha,y)   = 8\pi \int_{0}^{\pi} \mathrm{d} \xi \int_{0}^{\alpha} \mathrm{d} \eta  \frac{\sin^2 \xi  \sin \eta }{(1+\sin \xi \cos \eta )^3} \\
 \times \frac{\left(y^2 Z^2+\sin^2 \xi \sin^2 \eta -\cos^2 \xi \right)^2}{N(y,\xi,\eta)\left(1+\sin \xi \cos \eta \right)^2} .
\end{aligned}
\end{equation}
The derivation of these expressions can be found in the SI.

The various integrals can be solved explicitly for the cases $y=1$ and $y\rightarrow \infty$, but so far have eluded analytical evaluation for the general case $y\geq 1$ \cite{Kaznacheev2002,Prinsen2003}. Hence, we need to take recourse to a numerical evaluation and minimization with respect to the opening angle $\alpha$ and the bipolarness $y$. We recall that there is a one-to-one mapping between the opening angle $\alpha$ and the aspect ratio $x$ of the tactoids. From Eq.~\ref{eq5} we deduce that our parameter space is quite substantial: (i) the scaled volume of the droplets, $v$, (ii) the dimensionless anchoring strength $\omega$, (iii) the ratio of the bend and splay elastic constants $\kappa$, and (iv) the dimensionless strength of the magnetic field $\Gamma$.

Before numerically solving the pertinent equations in Section IV, we first analyze in Section III the problem from the perspective of scaling theory for (nearly) spherical and (highly) elongated droplets. This allows us to demarcate the crossovers between the various parameter regimes, and find the scaling exponents relevant to the behavior of the droplet shape and director field. As we shall see in Section IV, our scaling theory and our variational theory are amazingly consistent with each other. 

For those not interested in the full scaling analysis, which is rather technical in nature and details transitions between no fewer than five regimes, we refer to Figs.~\ref{fig3} and \ref{fig4_1} that summarize our main findings. The scaling relations that we find are summarized in Tables I, II and III, and Table IV lists all crossover volumes and external field strengths.

\section{Scaling Theory}
Rather than getting the exact expression for the free energy Eq.~\ref{eq5}, we may also estimate the equilibrium shape and director-field configuration of a tactoid by applying simple geometric arguments and or resorting to asymptotic relations valid for the various integrals that we introduced in the preceding section. We assume that the droplet looks like a spindle with the short axis $r$ and the long axis $R\geq r$, and that its director-field is quasi bipolar, {\it i.e.}, the director field converges outside the droplet to virtual point defect or boojums, see Fig.~\ref{fig2-1}. The (half) distance between the (virtual) defects is $\tilde{R}\geq R$.

Referring to the free energy functions Eqs.~\ref{TotalFreeEnergy}-\ref{ElectricFieldFreeEnergy}, we notice that the elastic and electric-field contributions must be proportional to the droplet volume $V \propto R r^2$, while the surface contributons must be proportional to the area of the droplet $S \propto rR$. Following Prinsen \textsl{et al.} \cite{Prinsen2004a}, we argue that the radius of curvature of a bend deformation must scale as $\tilde{R}^2/r$ and that of the splay as $\tilde{R}^2/R$.

Furthermore, the anchoring strength term, proportional to $(\vec{q} \cdot \vec{n})^2$, scales as $(r^2/R^2)(1-R^2/\tilde{R}^2)^2$ \cite{Prinsen2004a,jamali2017}, as we in fact also show in the SI.  Finally, the field term $(\vec{E}\cdot\vec{n})^2$ is proportional to $E^2 r^2 R^2 / \tilde{R}^4$. 

As already alluded to in the previous section, there are two quantities that a nematic droplet can optimize in order to lower its free energy: the aspect ratio $x=R/r\geq 1$ and the bipolarness $y = \tilde{R}/R\geq 1$. The dimensionless free energy of a droplet with aspect ratio $x$ and bipolarness $y$ reads within our scaling Ansatz
\begin{equation}
\begin{aligned}
f(x,y) \sim & \quad v^{2/3} x^{1/3} (1+\omega x^{-2} (1-y^{-2})^2)
\\
& +v^{1/3} y^{-4} x^{-4/3} (1+\kappa x^{-2}) +\Gamma v x^{-2} y^{-4},
\label{Free_Energyq_quasi}
\end{aligned}
\end{equation}
where we have ignored all constants of proportionality. The values of the dimensionless anchoring strength $\omega$, bend constant $\kappa = K_{33}/\left(K_{11}-K_{24}\right)$, electric field strength $\Gamma = \frac{1}{8\pi}\sigma ^{-2}\epsilon_a E^2 \left(K_{11}-K_{24}\right)$ and volume $v\equiv V \sigma^3/(K_{11}-K_{24})^3$ determine what values of $x$ and $y$ minimize the free energy $f$. The first term in Eq.~\ref{Free_Energyq_quasi} represents the surface free energy, the second term the elastic deformation and the last term the interaction of the droplet with the external electric field. Notice that the various terms can also be derived from Eq.~\ref{TotalFreeEnergy} by applying a formal expansion for small $\alpha \simeq x^{-1}$, and keeping only the leading order term of each contribution. We refer to SI for details. 

As we shall see, there are always two terms that dominate the shape and director field of a tactoid: either the surface and elastic energy, or the surface and Coulomb energy. This is a result of the different scaling with the dimensionless volume $v$: $v^{1/3}$ for the elastic free energy, $v^{2/3}$ for the interfacial free energy and $v$ for the Coulomb free energy. This means that droplet size crucially determines the shape and director field behavior of tactoids.

It is important to note at this point that the specific form of our scaling Ansatz for the free energy, Eq.~\ref{Free_Energyq_quasi}, automatically ensures that $y\geq 1$ but not that $x\geq 1$. The former follows directly from the observation that the first term has a minimum for $y=1$, whilst all the other terms decrease as $y>0$ increases. Further, Eq.~\ref{Free_Energyq_quasi} does \textsl{not} hold for lens-shaped tactoids, that is, for $x<1$. Indeed, the surface free energy for $x\ll 1$ would require a term proportional to an area that scales as $r^2$ rather than the $Rr$ that is valid for $x\geq 1$. Similar arguments hold for the elastic and Coulomb terms \cite{verhoeff2011,verhoeff2011tactoids,otten2012}.

In order to deal with the fact that our free energy does not automatically ensure the condition that $x\geq 1$ for $\omega \geq 0$, we have to separate the case of elongated droplets with $x\gg 1$ from that of spheroidal droplets with $x\approx 1$. For $\omega < 0$ the tactoids become lens-shaped with $x<1$. \cite{verhoeff2011,verhoeff2011tactoids,otten2012}.

The latter can be investigated by putting $x=1$ in the free energy, and not optimizing with respect to both $x$ and $y$, but only with respect to $y$. The crossover from elongated to spheroidal emerges automatically from our analysis, as we shall see. In what follows, we first analyze the simpler case for which the aspect ratio is close to unity, and next consider the case where the aspect ratio is significantly larger than unity. 

\subsection{Nearly spherical tactoids}
As we shall see in the next subsection, tactoids are always nearly spherical if the anchoring strength $\omega$ is about unity or smaller (see Eq.~\ref{InterfaceFreeEnergy}), irrespective of the value of the scaled volume $v$ and that of the scaled strength of the electric field $\Gamma$. If $\omega \gg 1$, then the droplets become spherical only for a range volumes that we will specify below, and then only if the field strength is below some critical value. 

Minimizing the free energy $f(1,y)$ for nearly spherical droplets with respect to the bipolarness $y$, we find for its optimal value
\begin{equation}
\begin{aligned}
y^2 \simeq 1 + \omega^{-1} \left(1+\kappa \right) v^{-1/3}  + \Gamma \omega^{-1} v^{1/3}.
\label{eq6}
\end{aligned}
\end{equation}
This expression immediately highlights the importance of the volume of the droplet. The director field is uniform, corresponding to $y\gg 1$, either if $\omega^{-1} (1+\kappa) v^{-1/3} \gg 1$ or $\Gamma \omega^{-1} v^{1/3} \gg 1$. In other words, if $v \ll v_{-} = \omega^{-3}(1+\kappa)^3$ or $v \gg v_{+} = \omega^3 \Gamma^{-3}$. For droplet volumes $v_{-} \leq v \leq v_{+}$, the director field is (quasi) bipolar, and $y\approx 1$. 

Thus, we find that there are potentially three different regimes and two critical volumes that dictate the behavior of the droplet. For $v \ll v_{-}$, we obtain the scaling relation
\begin{equation}\label{eq:yvsmall}
y \sim v^{-1/6} \omega^{-1/2}(1+\kappa)^{1/2},
\end{equation}
 whilst for $v\gg v_{+}$, we have 
\begin{equation}\label{eq:yvlarge}
y\sim v^{1/6}\Gamma^{1/2}\omega^{-1/2}.  
\end{equation}
Notice that the exponents of $-1/6$ for volumes $v \ll v_{-}$ and of $+1/6$ for $v\gg v_{+}$ are universal. A summary of these results is given in Table \ref{table1}.

We notice that as the electric field strength increases, $v_{+}$ decreases and the two critical volumes merge into one critical volume: $v_{-} = v_{+}$. This happens at a critical electric field $\Gamma_\mathrm{c} \simeq \omega^2  (1+\kappa)^{-1}$. If $\Gamma \geq \Gamma_\mathrm{c}$, the bipolarness $y$ is always greater than unity for any size of the droplet, and the droplet is never fully bipolar. The crossover of a decreasing bipolarness to an increasing one with increasing volume then happens at a critical volume $v_\mathrm{c} = \omega^{-3}(1+\kappa)^{3}$, where we inserted $\Gamma_c$ in the expression for $v_{+}$. If $\Gamma > \Gamma_c$, we find the crossover to occur for
$v_\mathrm{c} = \Gamma^{-3/2}(1+\kappa)^{3/2}$ that can be found by equating Eq.~\ref{eq:yvsmall} and \ref{eq:yvlarge}.  The summary of the results of this subsection is presented in Table \ref{table1}.

So, in conclusion, if $\Gamma > \Gamma_c$, then $y \gg 1$ decreases with increasing volume $v$ of the tactoid until larger than the critical value $v_c$. If larger than $v_c$, $y$ increases again with increasing volume and $y$ does not approach the value of unit, that is, the tactoid does not become bipolar. For $\Gamma < \Gamma_c$, the director field is bipolar if $v_{-} < v < v_{+}$, but not outside of this range of volumes. For $v<v_{-}$ the bipolarness $y$ decreases with increasing volume, whilst for $v>v_{+}$ it increases with increasing volume. All in all this demarcates three scaling regimes for the degree of bipolarness of the director field.

As we shall see next, for elongated tactoids the number of regimes increases to five.

\begin{table}[]
\begin{tabular}{c|c|c|c|c|}
\cline{2-5}
&$\hspace{0.6 cm}v < v_- \hspace{0.6 cm}$& \multicolumn{2}{c|}{$\hspace{0.1 cm} v_- < v < v_+  \hspace{0.1 cm}$}& $\hspace{0.6 cm}  v_+ < v \hspace{0.6 cm}$   \\ \hline
\multicolumn{1}{|c|}{$\Gamma \leq \Gamma_c $} & \multicolumn{1}{|c|}{$y \sim \omega^{-1/2} v^{-1/6}$} & \multicolumn{2}{c|}{$y \sim 1$}&      \\ \cline{1-1} \cline{3-4}
\multicolumn{1}{|c|}{$\hspace{0.3 cm}\Gamma_c \leq \Gamma \hspace{0.3 cm}$} & \multicolumn{2}{c|}{$ \quad \quad  \left(1+\kappa\right)^{1/2} \quad\quad \ \ \quad$} & \multicolumn{2}{c|}{$y  \sim  \ \Gamma^{1/2} \omega^{-1/2} v^{1/6}$} \\ \hline
\end{tabular}
\caption{This table summarises the various scaling regimes for the bipolarness $y$ of nearly spherical tactoids ($x \simeq 1$) in the presence and absence of an electric field, for large, intermediate and small droplet sizes $v$, relative to the crossover volumes $v_{-}$ and $v_{+}$. If the electric field is weak and $0 < \Gamma < \Gamma_c$, the bipolarness of the droplet has three different regimes. Under a strong field, $\Gamma>\Gamma_c$ there are two regimes.  Expressions for the crossover volumes $v_{-}$ and $v_{+}$ and the critical field strength $\Gamma_c$ are listed in table IV. Notice that for $\Gamma = 0$, $v_{+}\rightarrow \infty$ and there are strictly speaking only two regimes.
}
\label{table1}
\end{table}

\subsection{Elongated tactoids}
For elongated tactoids, matters become significantly more complex. To calculate the optimal values for the bipolarness $y$ and the aspect ratio $x$ for $x\gtrsim 1$, we need to minimize the free energy Eq.~\ref{Free_Energyq_quasi} with respect to both $x$ and $y$. This gives rise to the following set of coupled equations, 
\begin{equation}
\begin{aligned}
& y^4  = \omega x^{-2} \left(y^2 - 1 \right)^2  \\ &
 +v^{-1/3}  x^{-5/3} (1 +\kappa x^{-2})  + v^{1/3}  \Gamma  x^{-7/3} , 
\label{eq:y4}
\end{aligned}
\end{equation}
and 
\begin{equation}
\begin{aligned}
    y^2  & =  1 +
 \omega^{-1}  x^{1/3}  \left(1+\kappa x^{-2}\right) v^{-1/3} \\ & +   \Gamma \omega^{-1}  v^{1/3} x^{-1/3}.
\label{eq:y2}
\end{aligned}
\end{equation}
Inserting the last two terms of Eq.~\ref{eq:y2} in Eq.~\ref{eq:y4}, we find 
\begin{equation}
\frac{x^2}{\omega} =  \left( 1 - y^{-2} \right) + \left( 1 - y^{-2} \right)^2.
\label{eq:x}
\end{equation}
Inserting this back in Eq.~\ref{eq:y2} produces a non-linear equation entirely in terms of the quantity $y$. Unfortunately, we have not been able to solve this expression exactly. It can of course be solved numerically, but this would obviously defeat the purpose of the scaling theory. Fortunately, the governing equations can be solved asymptotically in a number of useful limiting cases that we will discuss next.

For instance, we have seen in the preceding section that for very large and very small droplet volumes the director-field must be nearly homogeneous, implying that $y \gg 1$. We note that large and small here refers to the critical volumes $v_{-}$ and $v_{+}$, introduced already in the preceding subsection for nearly spherical tactoids but that now will conform to slightly different expressions given below. Eq.~\ref{eq:x} tells us that 
if $y \rightarrow \infty$, the director field is uniform, and the aspect ratio is (apart from a multiplicative constant) equal to $\sqrt{\omega}$. This is consistent with the exact result $x = 2\omega^{1/2}$ obtained by means of the Wulff construction for $\omega \geq 1$ \cite{Prinsen2003}.

Inserting $x \simeq \omega^{1/2}$ in Eq.~\ref{eq:y4} gives, to leading order for large values of $y$,
\begin{equation}
y^2 \sim v^{-1/3}\omega^{-5/6} \left( 1 +\kappa \omega^{-1} \right) + v^{1/3}  \omega^{-7/6} \Gamma.
\label{eq:21}
\end{equation}
This means that for sufficiently small droplets
\begin{equation}
y \sim v^{-1/6}\omega^{-5/12}\left(1+\kappa \omega^{-1}\right)^{1/2},
\label{eq:yvsmall2}
\end{equation}
whilst if they large sufficiently large we have
\begin{equation}
y \sim v^{1/6}  \omega^{-7/12} \Gamma^{1/2}.
\label{eq:yvlarge2}
\end{equation}
It is worth mentioning that the result for large droplets does \textsl{not} depend on the value of $\kappa$ that is a measure for the magnitude of the bend elastic constant. 
Recall that in Eqs.~\ref{eq:yvsmall} and \ref{eq:yvlarge}, we find the same scaling of the bipolarness with the dimensionless volume for nearly spherical droplets. The scaling with the anchoring strength is slightly different, however. 

Equation~\ref{eq:yvsmall2} applies for $v\ll v_{-} = \omega^{-5/2}(1+\kappa \omega^{-1})^3$  and Eq.~\ref{eq:yvlarge2} for $v\gg v_{+} = \Gamma^{-3} \omega^{7/2}$, as can be deduced from Eq.~ \ref{eq:21}. These critical volumes differ slightly from those we calculated for nearly spherical tactoids, as already announced. 
We conclude that if $v_{-} \leq v \leq v_{+}$ the elongated droplets must be bipolar. 
 
If $v_{+}$ drops below $v_{-}$, the tactoids are always more or less uniform, and $x \simeq \omega^{1/2}$. This happens at a critical field strength $\Gamma_c = \omega^2(1+\kappa \omega^{-1})$ that we find by setting $v_{+}=v_{-}$. The crossover of decreasing to increasing bipolarness with increasing volume now occurs at a critical volume $v_c = \omega^{-3}$ for $\Gamma = \Gamma_c$. If $\Gamma > \Gamma_c$, the crossover happens at a critical volume $v_c = \Gamma^{-3/2} \omega^{1/2}(1+\kappa \omega^{-1})^{3/2}$, which we find by equating Eqs.~\ref{eq:yvsmall2} and \ref{eq:yvlarge2}.

What the aspect ratios of the tactoids are when $v_{-} \leq v \leq v_{+}$, so when the director field is not longer uniform, can be inferred from Eqs.~\ref{eq:y2} by inserting $y = 1+\delta$ in Eq.~\ref{eq:x} and presuming that $\delta \ll 1$. Solving these equations the gives to leading order in $\delta  = x^2 \omega^{-1} \ll 1$ an expression for the aspect ratio:  $x^{5/3} \sim v^{-1/3} (1+\kappa x^{-2}) + v^{1/3} \Gamma x^{-2/3}$. For small droplets with a volume $v_{+} \gg v >v_{-}$, we have $x \sim v^{-1/5}$ if we ignore the contribution from the bend elasticity; thus the droplet becomes less elongated with increasing volume. For larger ones, $v_{-}\ll v < v_{+}$, the aspect ratio $x \sim v^{1/7}\Gamma^{3/7}$ grows again with increasing volume. 

As we need to insist that $x>1$ for the equations to hold, we take the value of $x=1$ as the crossover to the regimes where the droplets are more or less spheroidal. Inserting this condition in the equation for the aspect ratio gives $1 \sim v^{-1/3} (1+\kappa ) + v^{1/3} \Gamma $, which we translate in two crossover volumes.
In the absence of a field, the crossover from an elongated droplet with $x=\omega^{1/2}$ to a nearly spherical droplet with $x = 1$ occurs for $v = v_{<} $ with $v_{<} \equiv (1 + \kappa)^3 > v_{-}$ another crossover volume. For sufficiently weak fields, they start to elongate again if $v_{>} < v < v_{+} = \Gamma^{-3} \omega^{7/2}$ with the crossover volume $v_{>} \equiv \Gamma^{-3}$ smaller than $v_{+}$ since $\omega >1$. For $v>v_{+}$ the director field is uniform and the aspect ratio obeys again $x \sim \omega^{1/2}$. 

\begin{figure}[h]
	\begin{center}
		\includegraphics[width=8cm]{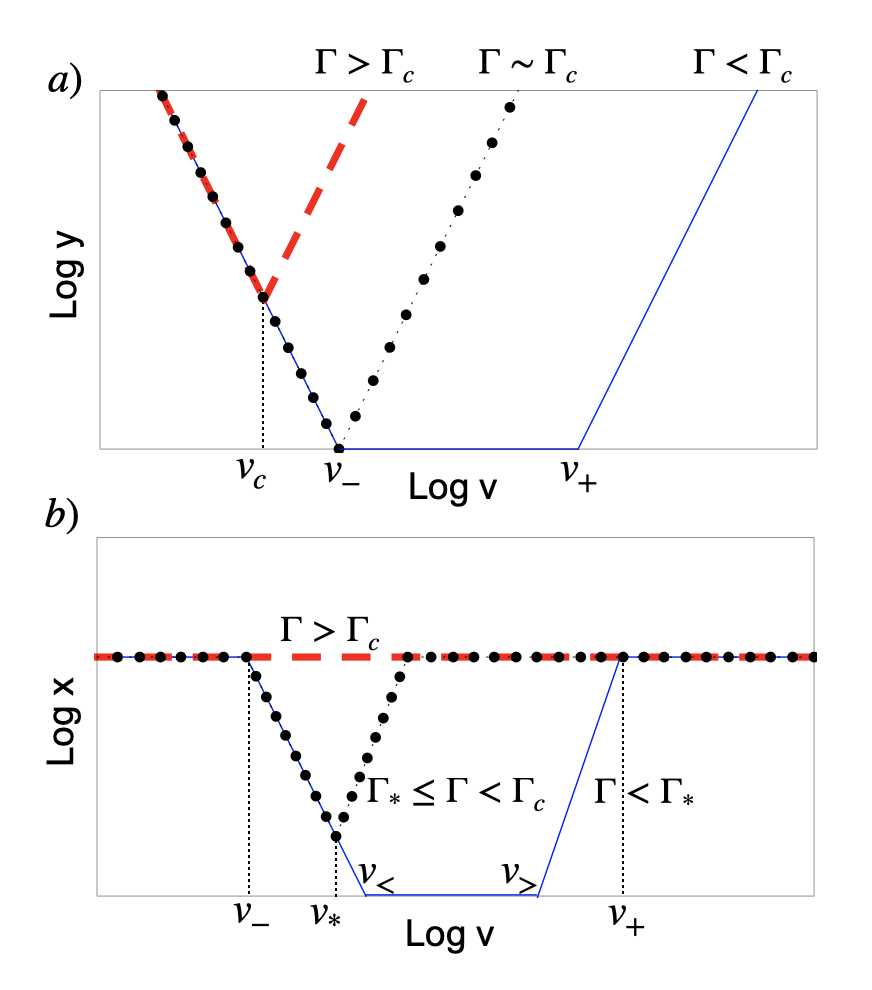}
		\caption{Dependence of (a) the degree of bipolarness $y$ and (b) the aspect ratio $x$ of tactoids on the scaled volume $v$ in the presence of an electric field, according to the scaling theory. Blue solid line: the field strength $\Gamma$ is below the critical value $\Gamma_c$, and the anchoring strength $\omega$ is somewhat larger than unity. The volumes $v_{-}$ and $v_{+}$ demarcate crossovers from quasi bipolar director fields to bipolar ones, and $v_{<}$ and $v_{>}$ those from elongated to spherical droplet shapes. Dash-dotted line: same curve, but now for $\Gamma=\Gamma_c$. Dashed lines: $\Gamma>\Gamma_c$. For fields above the critical field strength $\Gamma_*$, the tactoids are always elongated provided the anchoring strength $\omega$ is sufficiently large. See the main text. The slopes of the various curves are listed in Tables I, II and III, and the values of the various crossover volumes and field strengths in Table IV.}
		
		\centering 
		\label{fig3}
	\end{center}
\end{figure}

The picture that emerges is one where for $v < v_{-}$ the nematic droplets have a more or less uniform director field with an aspect ratio of about $\omega^{1/2}$, and the same for $v> v_{+}$. If $v>v_{-}$, the droplets become increasingly bipolar and the aspect ratio decreases with increasing volume. 
If the scaled volume $v$ increases further to get closer to $v_{+}$, the bipolar character of the director field diminishes again with increasing volume, while the aspect ratio increases to its maximum value of about $\omega^{1/2}$, see Fig.~\ref{fig3}.
Notice that we have presumed that $\omega \gg 1$; otherwise, we would not have $x \gg 1$. 

Somewhere in the size range $v_{-} \leq v \leq v_{+}$, the droplets actually become nearly spherical, in which case the theory of the preceding section applies. This happens in the range of volumes for which $v_{<} < v < v_{>}$. The nearly spherical drop regime disappears if $v_{<} = v_{>}$. Equating these critical volume shows that this occurs for field strengths $\Gamma$ larger than the critical value of  $\Gamma_* \equiv (1+\kappa)^{-1}$. In that case we only have crossover from decreasing elongated to increasing elongated at a crossover volume $v_* = \Gamma^{-5/4}$. Since $\Gamma_* < \Gamma_c$, we conclude that for $\omega \gg 1$ we lose the spherical tactoid regime before we lose the bipolar director field.

\begin{figure}[h]
	\begin{center}
		\includegraphics[width=8cm]{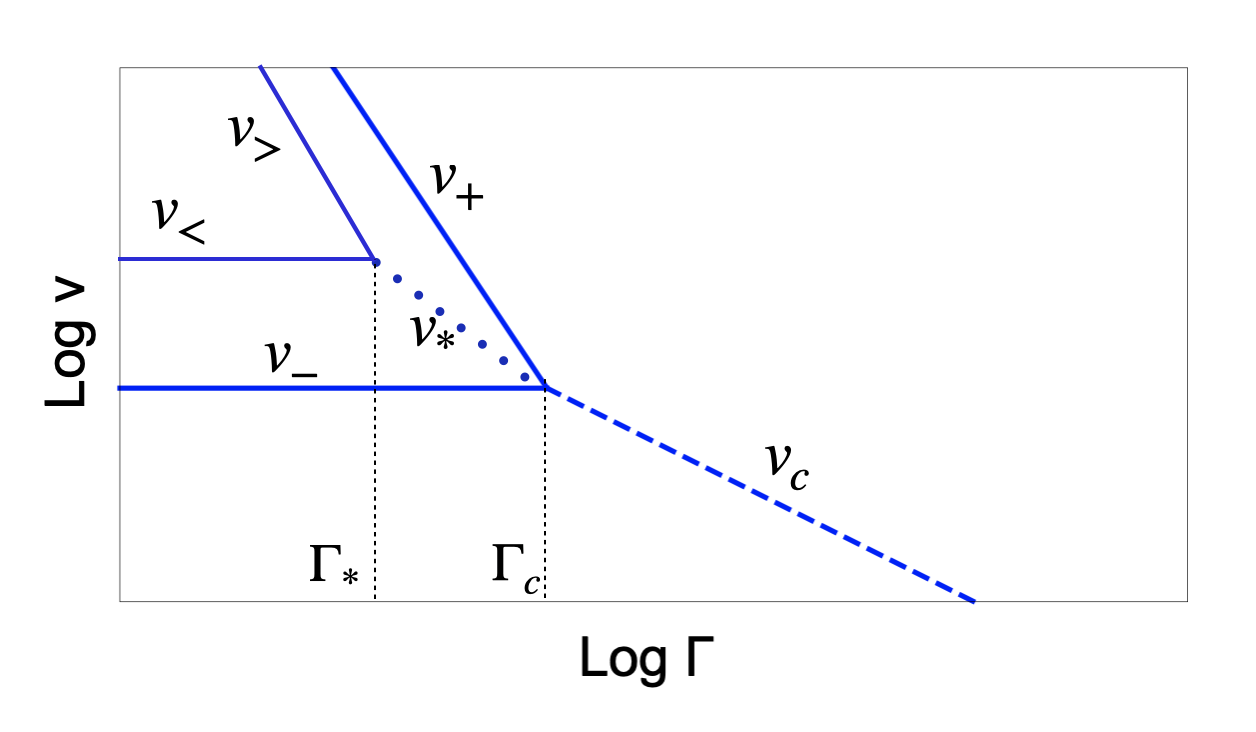}
		\caption{Schematic ``phase'' diagram of director fields of tactoids in an external field. Plotted is the scaled volume $v$ versus the scaled field strength $\Gamma$.  Indicated are the crossovers between the five regimes, demarcated by the crossover volumes $v_{-}$, $v_{+}$, $v_{<}$, $v_{>}$, $v_*$ and $v_c$. Expressions for the crossover volumes $v_{-}$, $v_{+}$, $v_{<}$ and $v_{>}$, and the critical field strengths $\Gamma_c$ and $\Gamma_*$, are listed in table IV. See also the main text. In the region bounded by $v_{-}$ and $v_c$, tactoids with sufficiently large anchoring strength $\omega \gg 1$ are elongated with a director field that is quasi bipolar with a bipolarness $y$ that decreases with increasing volume. In the region bounded by $v_{c}$ and $v_{+}$ in the upper right-hand corner it is quasi bipolar with a bipolarness that increases with droplet volume. In region bounded by $v_{-}$ and $v_{+}$ in the upper left-hand corner the director field is for all intents and purposes bipolar. The tactoids are more or less spherical in the region bounded by $v_{<}$ and $v_{>}$ in the upper most left-hand corner, and elongated outside of that region, at least if $\omega \gg 1$. The volume $v_c$ demarcates the crossover from decreasing to increasing bipolarness for quasi bipolar director fields, the volume $v_*$ that between decreasing aspect ratio to increasing aspect ratio.}	
		\centering 
		\label{fig4_1}
	\end{center}
\end{figure}

The summary of these results are presented in Tables \ref{table2}, \ref{table3}, and \ref{table4}, as well as in Fig.~\ref{fig4_1}, showing the different regimes and crossovers. The conclusion of our scaling theory is that the aspect ratio of the nematic droplets is at most $\omega^{1/2}$, independent of the volume or the field strength. In other words, external fields cannot elongate a tactoid to aspect ratios beyond those that are found in the absence of a field, at least for the family of director fields that we presume. We return to this issue in the Discussion section below.

In the next section, we present the numerical evaluation of our variational theory and obtain the mathematically exact response of the director field and the shape of the droplet in the presence of an electric field, and compare these with the scaling theory. As we shall see, our scaling predictions are robust. This means also that the conclusions that we base on them are robust.

\begin{table}[]
\begin{tabular}{c|c|c|c|c|}
\cline{2-5}
&$\hspace{0.4 cm}v < v_- \hspace{0.4 cm}$& \multicolumn{2}{c|}{$\hspace{0.2 cm} v_- < v < v_+ \hspace{0.2 cm}$}& $\hspace{0.4 cm}  v_+ < v \hspace{0.4 cm}$   \\ \hline
\multicolumn{1}{|c|}{$\hspace{0.2 cm}\Gamma \leq\Gamma_c \hspace{0.2 cm}$} & \multicolumn{1}{|c|}{$y \sim \omega^{-5/12} v^{-1/6}$} & \multicolumn{2}{c|}{$y \sim 1$}&      \\ \cline{1-1} \cline{3-4}
\multicolumn{1}{|c|}{$\Gamma_c \leq \Gamma$} & \multicolumn{2}{c|}{$ \quad  \left(1+\kappa\omega^{-1}\right)^{1/2} \quad \quad \quad ~ $} & \multicolumn{2}{c|}{$y  \sim  \ \Gamma^{1/2} \omega^{-7/12} v^{1/6}$} \\ \hline
\end{tabular}
\caption{This table summarises the various scaling regimes for the bipolarness $y$ of elongated tactoids ($x \gg 1$) in the presence and absence of an electric field, for large, intermediate and small droplet sizes $v$, relative to the crossover volumes $v_{-}$ and $v_{+}$. If the electric field is weak and $0 < \Gamma < \Gamma_c$, the bipolarness of the droplet has three different regimes. Under a strong field, $\Gamma>\Gamma_c$ there are two regimes. Expressions for the crossover volumes $v_{-}$ and $v_{+}$ and the critical field strength $\Gamma_c$ are listed in table IV. Notice that for $\Gamma = 0$, $v_{+}\rightarrow \infty$ and there are strictly speaking only two regimes.}
\label{table2}
\end{table}
\begin{table}[]
\resizebox{0.48\textwidth}{!}{
\begin{tabular}{c|c|c|c|c|c|c|}
\cline{2-7}
  & $v \! < \! v_- $ & $v_-\!< \! v \! < \! v_< $& \multicolumn{2}{c|}{$v_< \!< \!v \!< \!v_> $}& $ v_> \! < \!v \! < \!v_+ $& $v_+ \! < \! v$ \\ \hline
 \multicolumn{1}{|c|}{$ \Gamma \! < \! \Gamma_*$} &   & & \multicolumn{2}{c|}{$x \sim 1$} &  &     \\ \cline{1-1}\cline{4-5}
 \multicolumn{1}{|c|}{$\Gamma_* \! <  \! \Gamma < \! \Gamma_c$} & & \multicolumn{2}{c|}{$ ~ ~ \quad  x \sim  v^{-1/5} ~ ~ \quad$}        & \multicolumn{2}{c|}{$x \sim \Gamma^{3/7} v^{1/7}$} &       \\ \cline{1-1}\cline{3-6}
 \multicolumn{1}{|c|}{ $\Gamma_c \! < \! \Gamma $ }  & \multicolumn{6}{c|}{$x \sim \sqrt{\omega}$}  \\ \hline
\end{tabular}}
\caption{This table summarises the various scaling regimes for the aspect ratio $x$ of tactoids in the presence and absence of an electric field, for large, intermediate and small droplet sizes $v$, relative to the crossover volumes $v_{-}$ and $v_{+}$, and $v_{<}$ and $v_{>}$. For simplicity, we have dropped any dependence of the aspect ratio on $\kappa$. Expressions for the crossover volumes $v_{-}$, $v_{+}$, $v_{<}$ and $v_{>}$, and the critical field strengths $\Gamma_c$ and $\Gamma_*$, are listed in table IV. Notice that for $\Gamma = 0$, $v_{+}\rightarrow \infty$ and $v_{>} \rightarrow \infty$.}
\label{table3}
\end{table}
 
\begin{table}[]
\begin{tabular}{c|c|c|}
\cline{2-3}
& $\omega \lesssim 1$ & $\omega \gtrsim 1$ 
\\
\hline
\multicolumn{1}{|c|}{$\hspace{0.5 cm}v_-\hspace{0.5 cm}$}& $\omega^{-3}\left(1+\kappa\right)^3$& $\omega^{-5/2}\left(1+\kappa \omega^{-1}\right)^3$
\\
\hline
\multicolumn{1}{|c|}{$v_+$}& $\omega^{3}\Gamma^{-3}$ & $\omega^{7/2}\Gamma^{-3}$
\\
\hline
\multicolumn{1}{|c|}{$v_c$}& $\hspace{0.2 cm}\Gamma^{-3/2}\left(1+\kappa\right)^{3/2}\hspace{0.2 cm}$ & $\hspace{0.2 cm}\Gamma^{-3/2}\omega^{-1}\left(1+\kappa \omega^{-1}\right)^{3/2}\hspace{0.2 cm}$
\\
\hline
\multicolumn{1}{|c|}{$v_<$}& $-$&$\left(1+\kappa\right)^{3}$ 
\\
\hline
\multicolumn{1}{|c|}{$v_>$}& $-$& $\Gamma^{-3}$
\\
\hline
\multicolumn{1}{|c|}{$v_*$}& $-$&$\Gamma^{-5/4}$ 
\\
\hline
\multicolumn{1}{|c|}{$\Gamma_*$}& $-$ &$\left(1+\kappa\right)^{-1}$ 
\\
\hline
\multicolumn{1}{|c|}{$\Gamma_c$}&$\omega^2\left(1+\kappa\right)^{-1}$ & $\omega^2\left(1+\omega^{-1}\kappa\right)^{-1}$ 
\\
\hline

\end{tabular}
\caption{Listing of all crossover volumes $v$ and critical field strengths $\Gamma$, for small and large values of the anchoring strength $\omega$. Crossovers form elongated to spherical tactoids only occur if $\omega$ is sufficiently large.}
\label{table4}
\end{table}
\section{Numerical Results}
The scaling theory of the preceding section has enabled us to identify different scaling regimes, which we now investigate by numerically minimizing the free energy Eq.~\ref{eq5}. 
To this end, we evaluate Eq.~\ref{eq5} for opening angles $0\leq \alpha \leq \pi$ and degrees of bipolarness of the director field $1\leq y\leq \infty$ to find the values of these quantities for which the free energy $f$ is minimal. So, for a given scaled volume $v$, anchoring strength $\omega$, ratio of bend-to-splay elastic constants $\kappa$ and electric field strength $\Gamma$, we obtain the optimal values of both $\alpha$ and $y$. We recall that the aspect ratio of the tactoids $x$ is directly linked to the opening angle via the relation $x= \cot (\alpha/2)$. In order to find the minimum free energy, we numerically calculate all integrals given in Sect. II. 
 
It is clear that the electric field drives the director field to align itself with it, implying that the major axis of a tactoid orients parallel to the electric field. This happens irrespective of whether the director-field configuration is uniform or bipolar. If the electric field is sufficiently weak, the director-field is not perturbed by the electric field. If the field is sufficiently strong, we would expect the director field of the droplet to become homogeneous, even if the director field in the absence of a field is bipolar. What weak and strong here mean depends on volume of a tactoid as we have seen in the preceding section, and is schematically summarized in the Figs.~\ref{fig3} and \ref{fig4_1}.

\begin{figure}[h]
	\begin{center}
		\includegraphics[width=8cm]{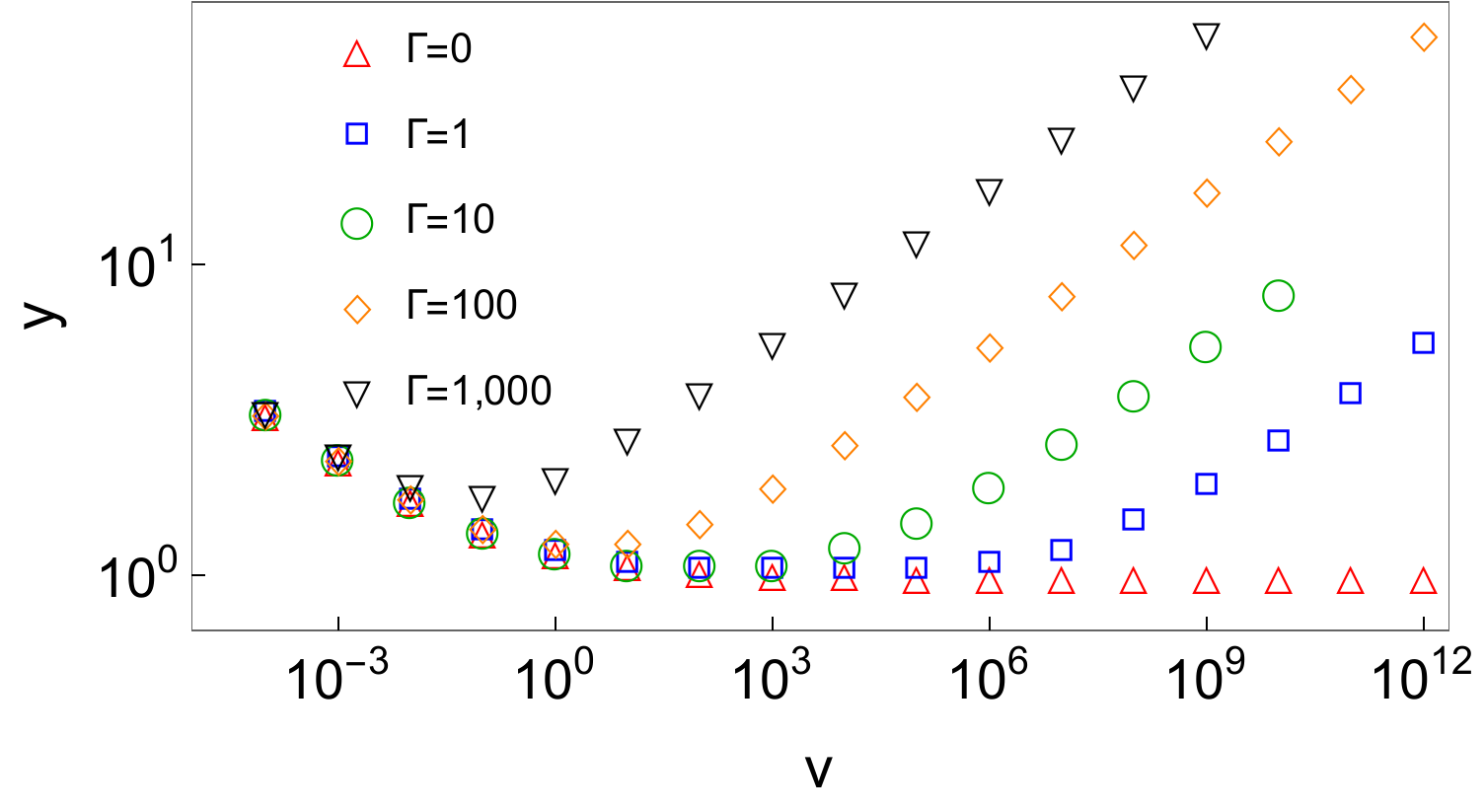}
		\caption{The bipolarness $y$ of the director field of a tactoid in the presence of an electric field as a function of its dimensionless volume $v$. Indicated  by the different symbols are results for different values of the dimensionless electric field strength $\Gamma$. The dimensionless anchoring strength is fixed at $\omega = 14$ and the dimensionless bend constant at $\kappa =10$. }	
		\centering 
		\label{fig4}
	\end{center}
\end{figure} 

Figure~\ref{fig4} confirms this expectation. Shown is the bipolarness $y$ as a function of the dimensionless volume $v$ of the tactoids, for the case where we (arbitrarily) set for the dimensionless bend constant $\kappa = 10$ and for the anchoring strength $\omega = 14$. Indicated are results for different values of the dimensionless electric field $\Gamma$. We confirm the scaling prediction that two critical volumes emerge, one associated with the crossover from a quasi bipolar to a bipolar director field, $v_{-}$, and one with the crossover from a bipolar to quasi bipolar director field, $v_{+}$. For $\Gamma \geq 100$, we only find quasi bipolar director fields characterized by a bipolarness $y>1$ for all volumes $v$. The scaling exponent $\beta$ we find for $y \sim v^\beta$ equals $\beta = -0.15$ for small volumes and $\beta = +0.16$ for large volumes, values that agree reasonably well with the predicted exponents of $-1/6$ and $+1/6$ that we obtained from the scaling theory and are quoted in table \ref{table2}.

\begin{figure}[h]
	\begin{center}
		\includegraphics[width=8cm]{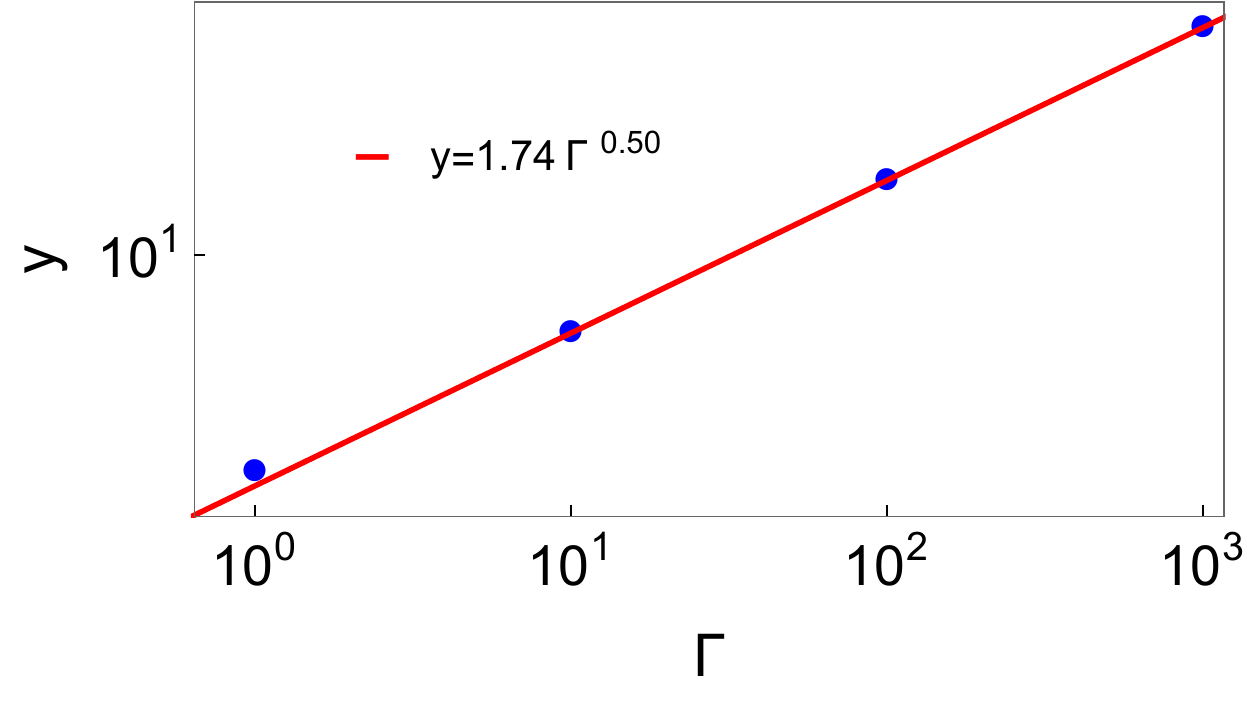}
		\caption{Bipolarness $y$ of a tactiods as a function of the dimensionless electric field $\Gamma$ for a dimensionless volume $v=10^7$. Anchoring strength $\omega =14$ and dimensionless bend constant $\kappa = 10$. The solid line shows the scaling of $y$ as with $\Gamma^{0.5}$. }	
		\centering 
		\label{fig5}
	\end{center}
\end{figure}

Figure~\ref{fig4} also shows that the bipolarness $y$ increases with the electric field strength $\Gamma$, if the volume of a tactoid is sufficiently large, $v>v_{+}$. According to the scaling prediction Eq.~\ref{eq:yvlarge2}, $y$ should scale as $\Gamma^{1/2}$. Figure \ref{fig5}, in which we plotted the bipolarness as a function of the field strength for the case where $\omega = 14$ and $\kappa =10$, confirms that the scaling exponent is $0.5$ over three decades of $\Gamma$. So, indeed, increasing the field strength leads to director fields that become increasingly homogeneous, as one would in fact expect from the scaling theory of the previous section. See also Table \ref{table2}. 

\begin{figure}[h]
	\begin{center}
		\includegraphics[width=8cm]{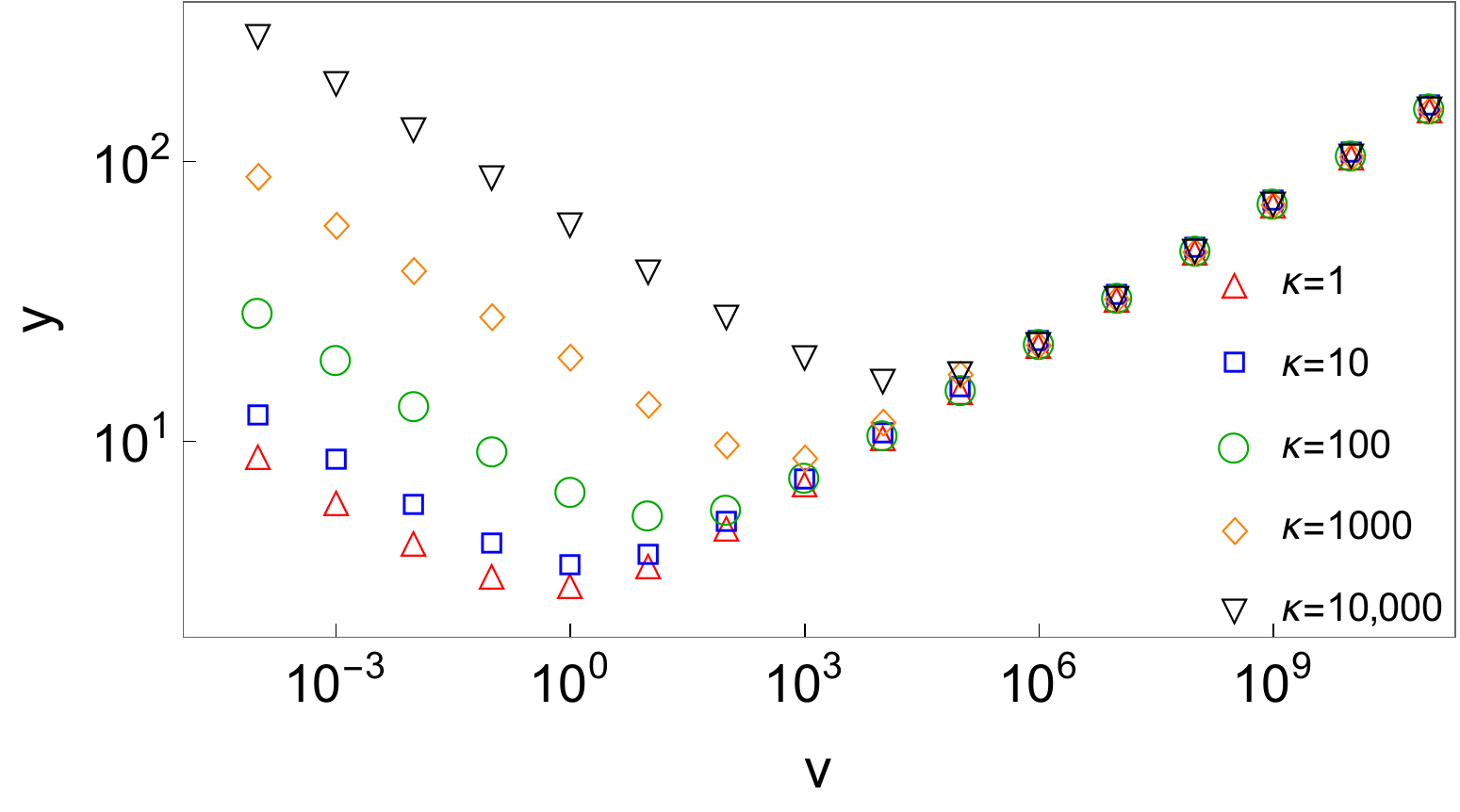}
		\caption{Bipolarness $y$ of the droplet as a function of the dimensionless volume $v$ of tactoids in the presence of an electric field for different values of the dimensionless bend constants $\kappa$ indicated by the symbols. Anchoring strength $\omega =14$ and dimensionless field strength $\Gamma =100$.}	
		\centering 
		\label{fig6}
	\end{center}
\end{figure}

According to the scaling theory of the preceding section, the impact of the (scaled) bend elastic constant $\kappa$ on the bipolarness $y$ of a tactoid is negligible for sufficiently large tactoids in the presence of an external field. See Table II. It is neglible for small tactoids too, but only provided $\kappa \ll \omega$. Our numerical results presented in Fig.~\ref{fig6} confirms for the case $\omega = 14$, the bipolarness is an invariant of $\kappa$ for sufficiently large volumes, but becomes a function of $\kappa$ for values larger than about 10, as expected from the scaling theory. 

\begin{figure}[h]
	\begin{center}
		\includegraphics[width=8cm]{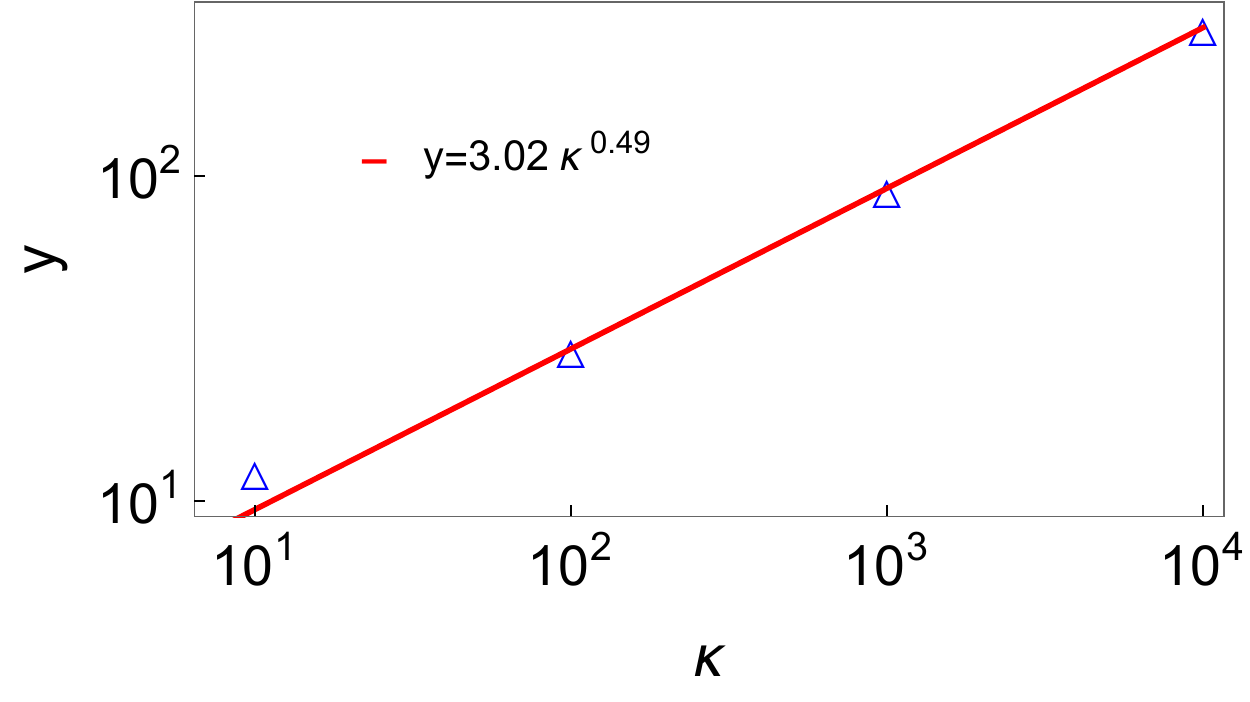}
		\caption{Bipolarness of a tactoid as a function of the dimensionless bend constant $\kappa$ in the presence of an electric field. Anchoring strength $\omega =1.4$, dimensionless field strength $\Gamma =100$ and dimensionless volume $v=10^{-4}$.}	
		\centering 
		\label{fig7}
	\end{center}
\end{figure}

For bend elastic constants $\kappa > 10 \approx \omega$, the bipolarness should exhibit a power-law scaling predicted by the scaling relation Eq.~\ref{eq:yvsmall2} that then takes the simpler form $y \sim v^{-1/6}\omega^{-11/12}\kappa^{1/2}$. In Fig.~\ref{fig7} we have plotted the bipolarness $y$ as a function of $\kappa$ for $\Gamma =100$ and $v=10^{-4}$. The exponent that we measure is $0.49$, which is indeed close to the value obtained from the scaling theory.

\begin{figure}[h]
	\begin{center}
		\includegraphics[width=8cm]{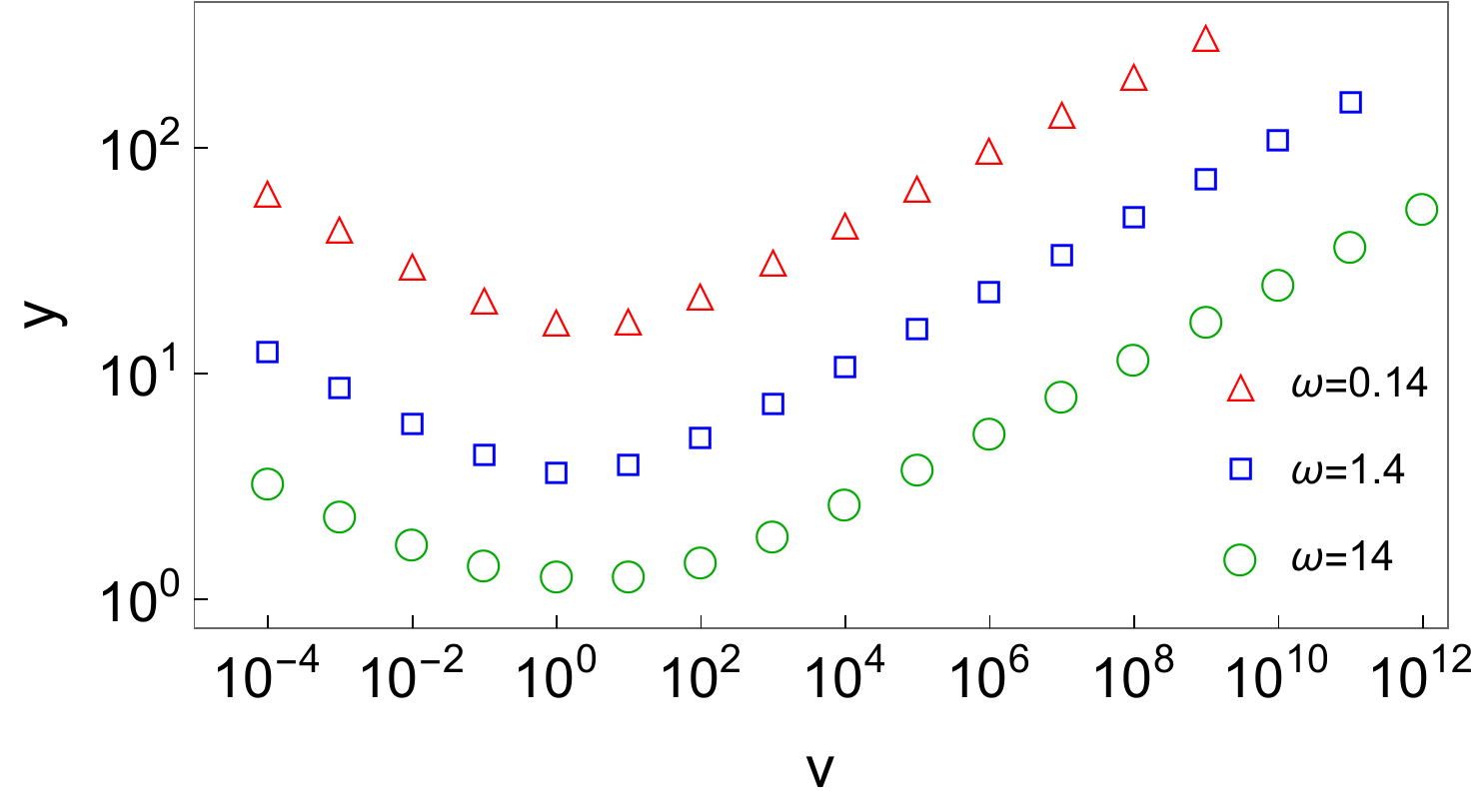}
		\caption{Bipolarness $y$ of a tactoid as a function of the dimensionless volume $v$ in the presence of an electric field for different values of the anchoring strength, indicated by the symbols. Dimensionless field strength $\Gamma =100$ and bend constant $\kappa = 10$.}	
		\centering 
		\label{fig8}
	\end{center}
\end{figure}

Our scaling theory also predicts the bipolarness of the tactoids to depend on the anchoring strength, $\omega$. Indeed, Eqs.~\ref{eq:yvsmall} and \ref{eq:yvlarge} for nearly spherical tactoids, and Eqs.~\ref{eq:yvsmall2} and \ref{eq:yvlarge2} for elongated ones, predict that both for small and large droplets the bipolarness should shift with shifting anchoring strength. This makes intuitive sense, because the larger the anchoring strength is, the larger the free energy penalty becomes for imperfect planar anchoring. Hence, with increasing anchoring strength the tactoids should become increasingly bipolar. This is what our numerical calculations also confirm, as is shown in Fig.~\ref{fig8}. On a logarithmic scale the curves shift vertically by an amount that depends on the anchoring strength $\omega$.

\begin{figure}[h]
	\begin{center}
		\includegraphics[width=8cm]{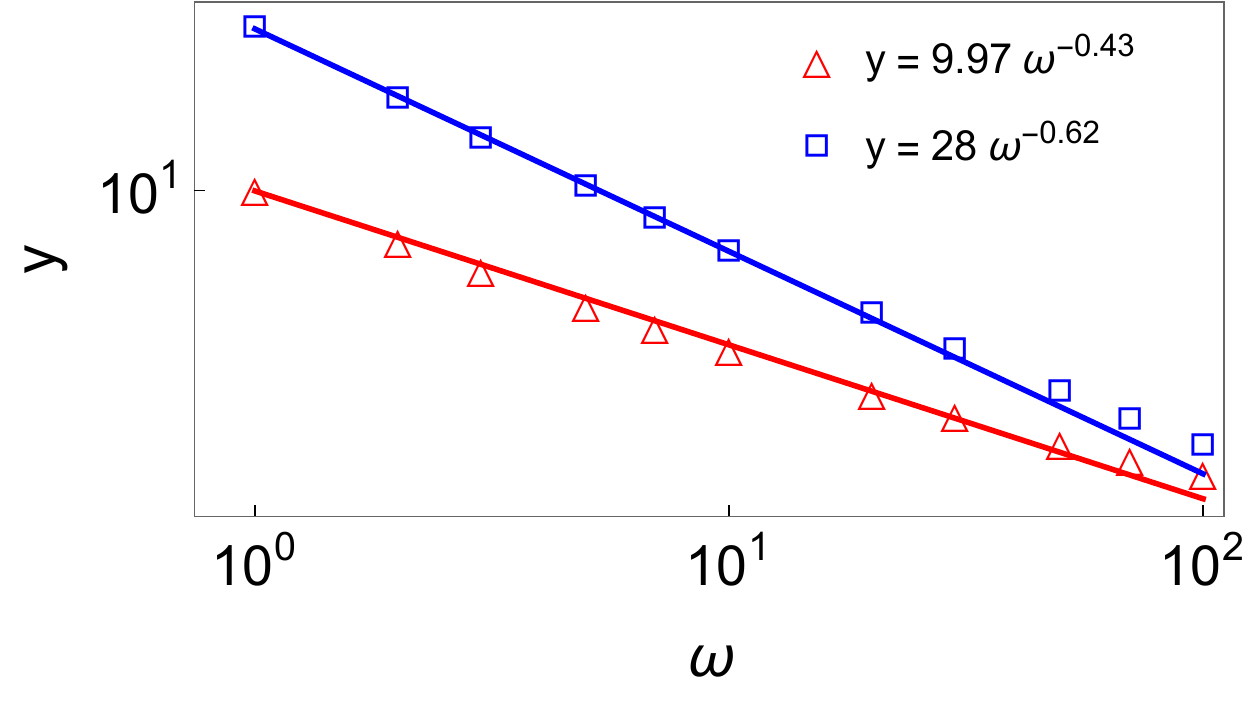}
		\caption{Bipolarness $y$ of a tactoid as a function of an anchoring strength $\omega>1$ for small and large droplets, with dimensionless volumes $v=10^{-2}$ and $v=10^6$. The dimensionless electric field strength is fixed at $\Gamma=10$ and the dimensionless bend constant at $ \kappa = 0$. Indicated are also the scaling relations $y\sim \omega^{-0.43}$ for the small volume and $y\sim\omega^{-0.62}$ for the large volume. See also the main text.}	
		\centering 
		\label{fig9}
	\end{center}
\end{figure}

\begin{figure}[h]
	\begin{center}
		\includegraphics[width=8cm]{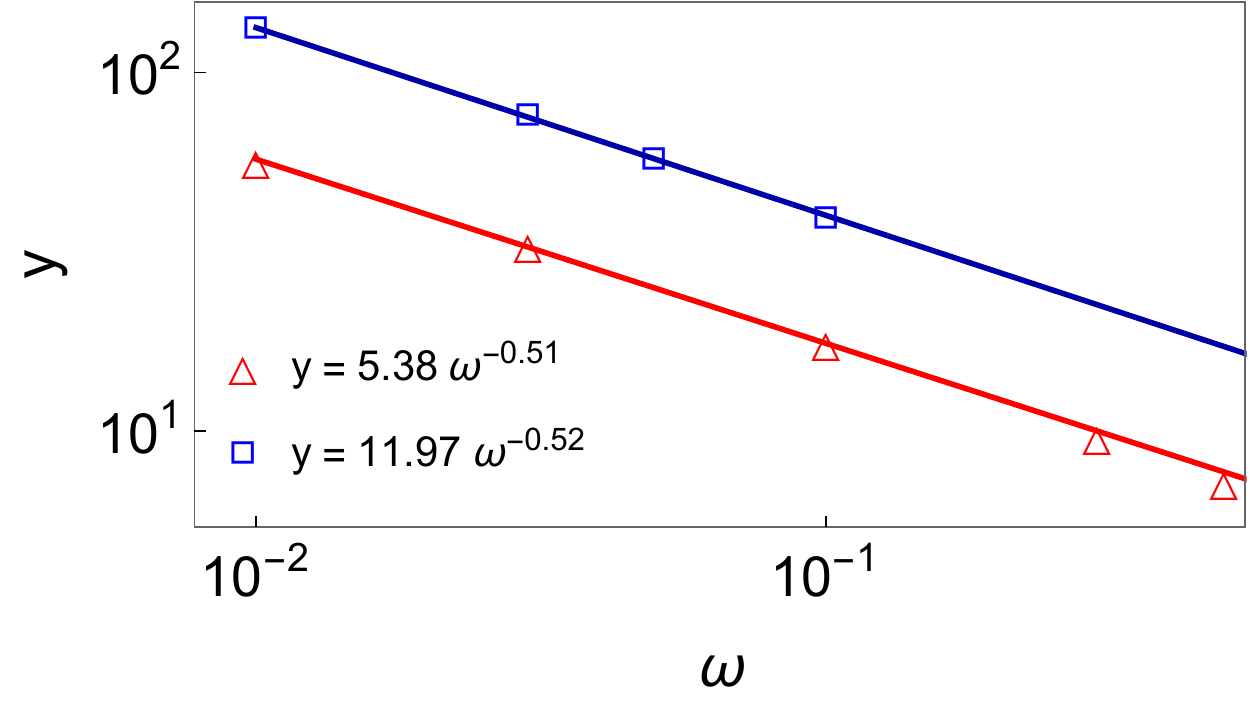}
		\caption{Bipolarness $y$ of a tactoid as a function of anchoring strength $\left(\omega<1\right)$ for small and large droplets, with dimensionless volumes $v=10^{-2}$ and $v=10^6$. The dimensionless electric field strength is fixed at $\Gamma=10$ and the dimensionless bend constant at $ \kappa = 0$. Indicated are also the scaling relations $y\sim \omega^{-0.51}$ and $y\sim\omega^{-0.52}$. See also the main text.}	
		\centering 
		\label{fig9_1}
	\end{center}
\end{figure}

The scaling of the bipolarness $y$ with the anchoring strength $\omega$ is highly non-trivial, as is implicit in the scaling predictions Eqs.~\ref{eq:yvsmall} and \ref{eq:yvlarge} for nearly spherical tacoids, and Eqs.~\ref{eq:yvsmall2} and \ref{eq:yvlarge2} for elongated ones. It depends not only on the shape of the tactoids, but also whether the tactoids are large or small, and on whether or not the bend elastic constant is large. To account for this, we plot in Figs.~\ref{fig9} and \ref{fig9_1} the bipolarness $y$ as a function of the anchoring strength $\omega$ for two droplet sizes and fixed values of $\Gamma=10$ and $\kappa=0$. The appropriate scaling regimes for $\omega <1$ for which the droplets are approximately spherical, and $\omega > 1$ for which they are elongated, are also illustrated in the figure.  Four different scaling exponents, which agree rather well with the predictions from scaling theory are shown in the figure.

\begin{figure}[h]
	\begin{center}
		\includegraphics[width=8cm]{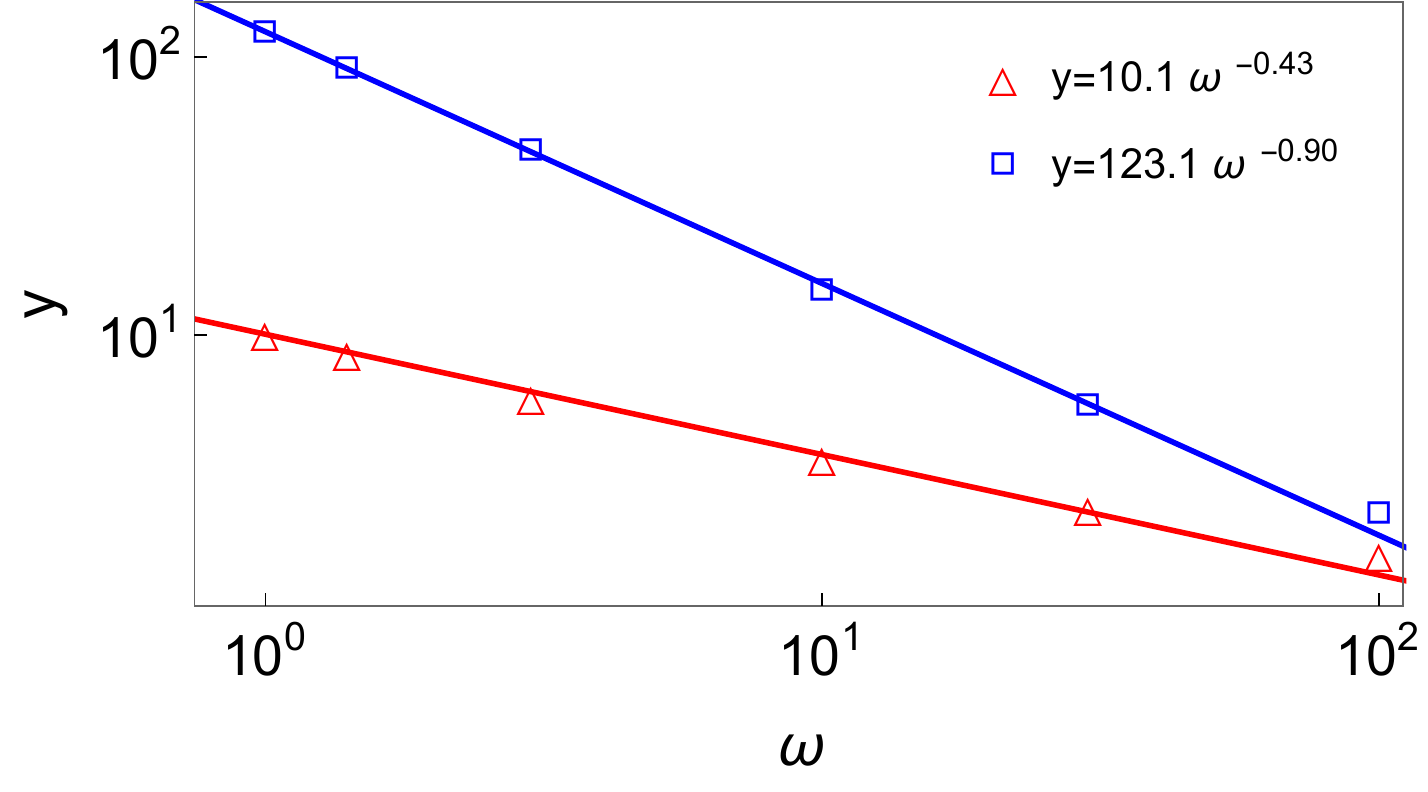}
		\caption{Bipolarness $y$ as a function of anchoring strength $\omega$ for two different  dimensionless bend constants of $\kappa =0.1$ and $\kappa =10^3$. The dimensionless volume is set at a value $v=10^{-4}$. Also indicated are the scaling exponent of $-0.43$ for the small value of the bend constant and of $-0.90$ for the large value of the constant. See also the main text. }	
		\centering 
		\label{fig10}
	\end{center}
\end{figure}

How the value of the bend elastic constant $\kappa$ impacts the dependence of the bipolarness $y$ and the anchoring strength $\omega$ is highlighted in Fig.~\ref{fig10} for a large and small value of $\kappa$. For the range of anchoring strengths shown, a small tactoid volume of $v = 10^{-4} $ and a field strength of $\Gamma = 10^2$, we find scaling exponents of $-0.43$ and $-0.90$ for the small and large values of dimensionless bend constants $\kappa$, which have to be compared with the scaling predictions of $-5/12\simeq -0.42$ and $-11/12\simeq -0.92$. Again, we find quite good agreement between our numerical work and the scaling theory. (See also table II.)

\begin{figure}[h]
	\begin{center}
		\includegraphics[width=8cm]{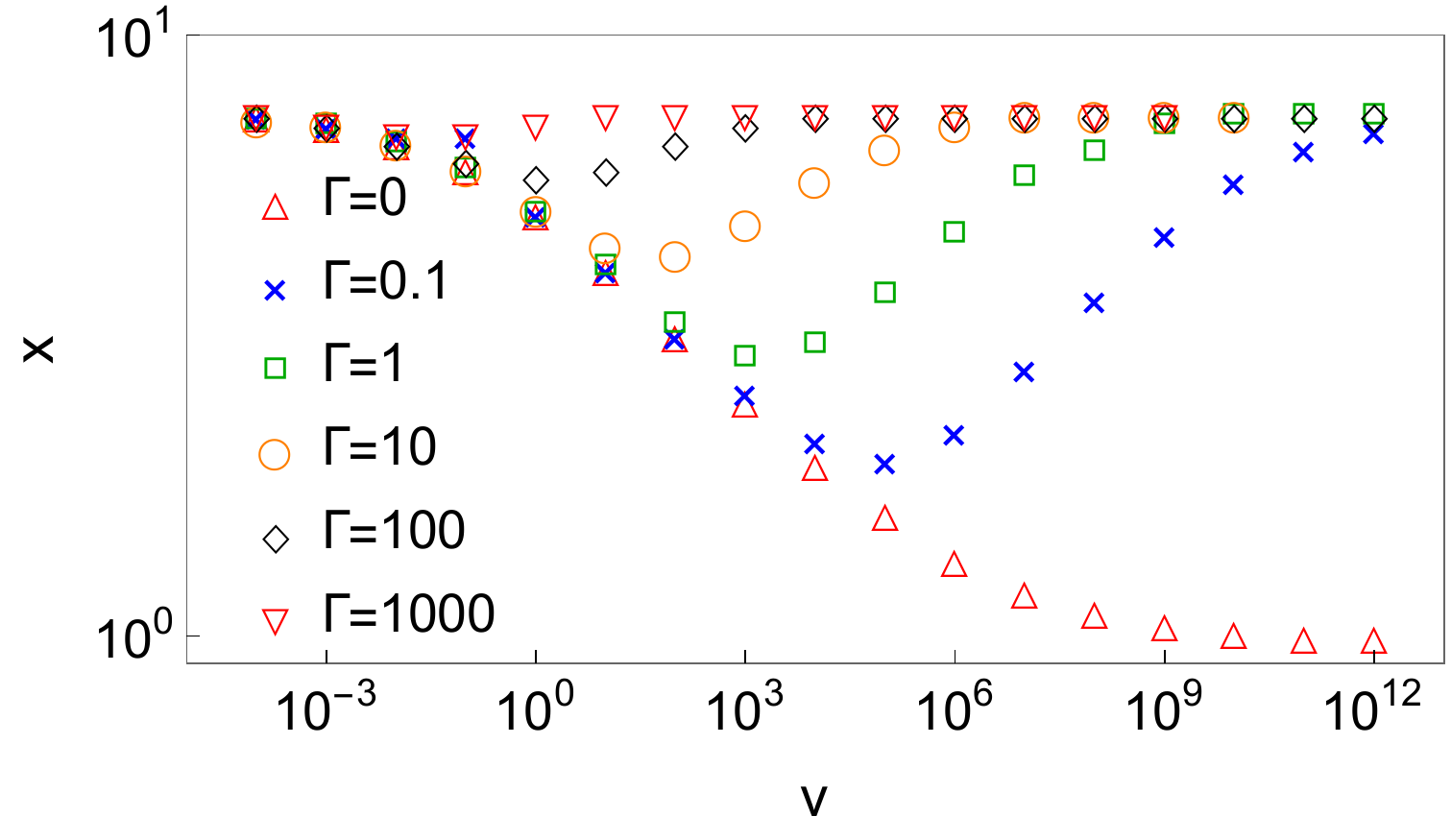}
		\caption{The aspect ratio $x$ of a tactoid as a function of the dimensionless volume $v$ for different (dimensionless) electric field strengths $\Gamma$ indicated by the symbols. The anchoring strength $ \omega =14 $ and the dimensionless bend constant $ \kappa=10$. }	
		\centering 
		\label{fig11}
	\end{center}
\end{figure}

Having exhaustively verified the theoretical scaling predictions for the degree of bipolarness of the tactoids, we now proceed to investigate how their aspect ratio depends on the volume and how it responds to the presence of an electric field. It is well known that, in the absence of an electric field, the aspect ratio of a nematic tactoid decreases with increasing droplet size. This happens to be so not only in bulk, but also if the tactoids deposited on a partially wetting surface \cite{Jamali2015,jamali2017,Metselaar2017}. Indeed, from the scaling theory we expect that for $v>v_{-}$, the aspect ratio $x$ should scale as $v^{-1/5}$ at least if $\kappa \ll \omega$ and $\omega \rightarrow \infty$ \cite{Prinsen2003,Prinsen2004a,jamali2017}. For finite $\omega =14$ the decay of the aspect ratio with volume is even a weaker function of the volume, as Fig.~\ref{fig11} shows for the field-free case $\Gamma = 0$.

Notice that for the dimensionless bend constant of $ \kappa=10$, the predicted critical magnetic field strength of $\Gamma_* = 1/11 \simeq 0.09$ coincides with the smallest non-zero value of $\Gamma$ taken in our numerical calculations. This means that our results of Fig.~\ref{fig11} should show conditions characterised by an absence of an intermediate regime with spherical tactoids, excluding the case $\Gamma =0$. See also the phase diagram of Fig.~\ref{fig4_1}. The predicted crossover volume $v_* \sim \Gamma^{-5/4}$ from a decreasing aspect ratio to an increasing aspect ratio varies $5$ orders of magnitude for the range of field strengths shown in the figure, in agreement with our numerical results presented in the figure.

What Fig.~\ref{fig11} also shows is that for increasingly large fields, the drop in aspect ratio becomes small mirroring the prediction of our scaling theory. This happens for $\Gamma > \Gamma_c \sim \omega^2/(1+\kappa/\omega) \simeq 100$ for our choice of parameters, 
when the drop in aspect ratio in fact disappears. This value is consistent with our findings. In that case the director field is for all intents and purposes uniform irrespective of the volume of the nematic droplet. This means that in our model, there is an upper limit for the aspect ratio, namely $2\sqrt{\omega}$. 

All of this implies that for our choice of director field geometry, an externally applied electric (or magnetic) field \textsl{cannot} elongate tactoids beyond their maximum aspect ratio that under the field-free conditions happens for sufficiently small droplets. This, clearly, goes against the grain of the experimental observations of Metselaar {\it et al.} of tactoids in electric fields \cite{Metselaar2017}, and those of Kaznacheev and collaborators in magnetic fields \cite{Kaznacheev2002}. As we argue in the next section, this must means that either (i) the director field does not conform to a bispherical geometry in an external alignment field; (ii) the tactoids are in a restricted equilibrium characterized by a bipolarness that is fixed to the value of the field-free initial state; or (iii) the various elastic and surface constants do depend on the strength of the field.

\section{Discussion and Conclusions}
In this paper, we present a model in which the director field and shape of a nematic tactoid can adjust themselves both in order to optimize the interfacial, elastic and Coulomb energy in the presence of an externally applied orienting field. We restrict the shape of the tactoid to that of the family of circle sections of revolution, and the director field to that of the family of fields that can be described by bispherical geometries. We find that the known "phase" diagram of nematic tactoids becomes more complex in the presence of an electric field. \cite{Prinsen2003,Prinsen2004a,Prinsen2004b} 

In the absence of such an alignment field there are three regimes, separating elongated tactoids with a uniform director field if they are sufficiently small from roundish bipolar ones if very large, with an intermediate size range where the drops are quasi bipolar and somewhat elongated. In the presence of an alignment field, we have identified up to five regimes depending on the strength of the anchoring of the director to the interface. A schematic of the new phase diagram is given in Fig.~\ref{fig4_1}.

Close comparison of theoretical predictions based on this model and experimental observations on tactoids of carbon nanotubes in chlorosulfonic acid by Jamali \textsl{et al.} have shown that, in the absence of an electric orienting field, there is a very good agreement between the theory and experiments \cite{Jamali2015}. The predicted gradual crossover from elongated to more or less spherical shapes, and from uniform to bipolar director fields, is confirmed experimentally, not only for tactoids in bulk solution but also for sessile tactoids, \textsl{i.e.}, tactoids on surfaces \cite{jamali2017}. Curve fits provide access to information on the surface energies and bend constants \cite{Kaznacheev2003,Prinsen2004a,Jamali2015,jamali2017,bagnani2019six,wang2018liquid,wang2018liquid,nystrom2018liquid,bharadwaj2000mesoscale,debenedictis2016shape}.

\begin{figure}[h]
	\begin{center}
		\includegraphics[width=8cm]{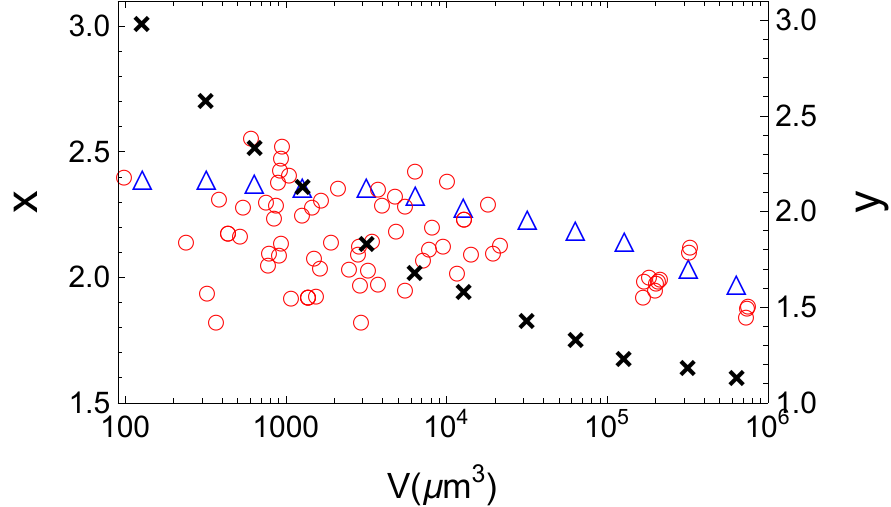}
		\caption{Aspect ratio $x$ and bipolarness $y$ of the director-field as a function of volume of a droplet in the absence of the electric field with $\Gamma=0$. Left vertical axis shows us the aspect ratio (we use blue triangles for the aspect ratio) and the right vertical axis shows us bipolarness (cross sign for bipolarness) and the red circles represent the experimental data of \cite{Metselaar2017}. The best fit by eye we obtain, using the parameter values $\kappa = 20$, $\omega = 1.3$ and $\left(K_{11}-K_{24}\right)/\sigma = 4 $ $\mu$m. Notice that the largest tactoids are bipolar because $y\rightarrow 1$ and the smalles ones quasi bipolar with $y \approx 3$.}
		\label{fig13}
	\end{center}
\end{figure}

For instance, if we curve fit the theory to the experimental data of Metselaar \textsl{et al.} on tactoids formed in dispersions of chitin fibres in water in the absence of an electric field, we obtain a reasonably good agreement if we set $\omega = 1.6$, $\kappa = 20$  and $(K_{11}-K_{24})/\sigma = 4$ $\mu$m. See Fig. \ref{fig13}, where the aspect ratio $x$ is plotted against the actual volume of the droplets. Also shown in the figure is the predicted bipolarness $y$ of the tactoids, which vary between $3$ and just over $1$ over that range of droplet volumes. It suggests that the tactoids of chitin in water are either bipolar or quasi bipolar, in agreement with experimental observation. \cite{Metselaar2017}

Because of the scatter in the data, and since we do not cover the whole range of volumes from nearly uniform to bipolar director fields as was done in the work of Jamali \textsl{et al.} \cite{Jamali2015}, we cannot expect these estimates to be highly accurate. 
Still, if we take them at face value, we find them to differ quite substantially form the ones found by Jamali \textsl{et al.} for carbon nanotubes in chlorosulfonic acid, with $\omega = 5.6$, $\kappa = 1.3$ and $(K_{11}-K_{24})/\sigma = 78$ $\mu$m \cite{Jamali2015}. This, however, should not be too surprising, given that both the elastic constants and surface energies depend sensitively on the dimensions of the particles \cite{sato1996,chen2002,van1999}.

Rather unexpectedly, our predictions fail if an external field is applied. In the experiments of Metselaar \textsl{et al.}, sufficiently large tactoids elongate up to ten times their original aspect ratio, which is much more elongated than the droplets in the absence of a field \cite{Metselaar2017}. As we have seen in our model calculations, the presence of very large field strengths do \textsl{not} lead to highly elongated shapes but to uniform director fields. As already announced, this might perhaps suggest that the bipolarness of the tactoids cannot respond sufficiently swiftly to the switching on of the external field. Before discussing the accuracy of this presumption, we first investigate its consequences assuming that it is true. 

The procedure that we pursue is as follows. First we calculate the bipolarness $y$ of the director field for the field-free case with $\Gamma=0$. Next, in the presence of an orienting field, so for $\Gamma>0$, we use this value of the bipolarness and optimise the free energy only with respect to the aspect ratio $x$. 
 
\begin{figure}[h]
	\begin{center}
		\includegraphics[width=8cm]{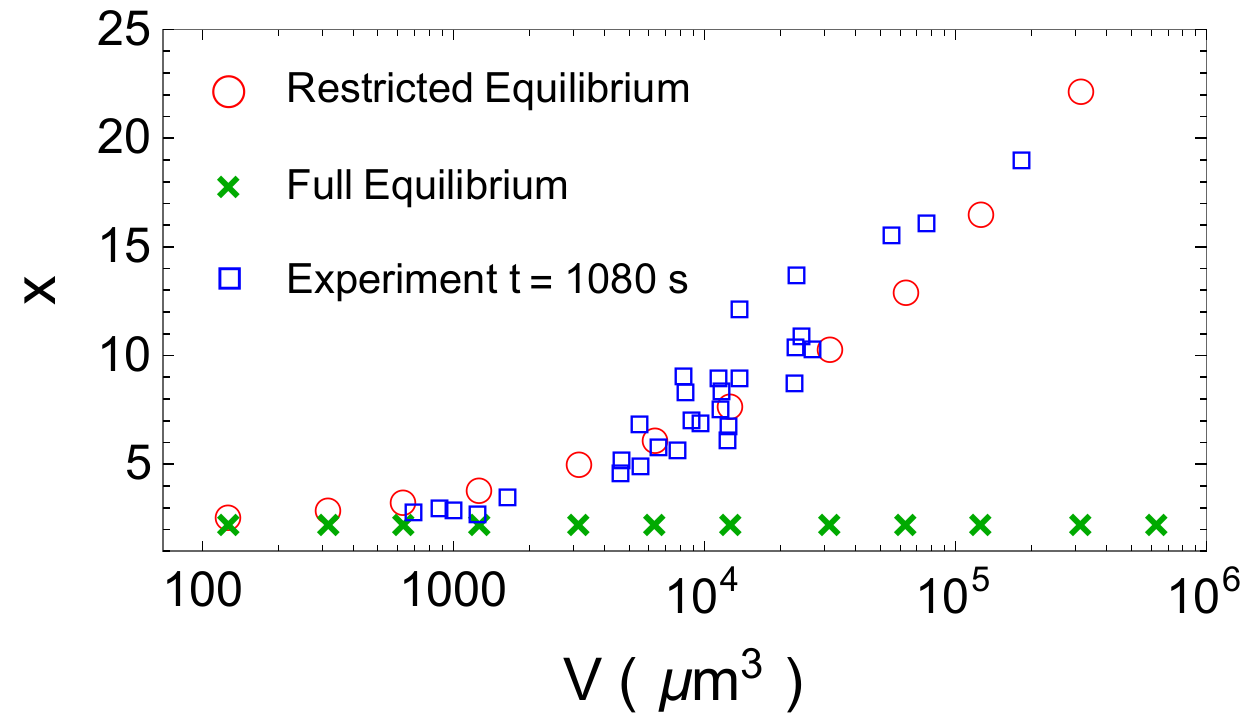}
		\caption{Aspect ratio $x$ of tactoids as a function of volume $V$ in the presence of an electric field. We compare our numerical results with the experimental data of \cite{Metselaar2017}. The best fit by eye we obtain using the parameter values $\Gamma = 500$, $\omega = 1.3$, $\kappa=20$ and $\left(K_{11}-K_{24}\right)/\sigma = 4 $ $\mu$m.}	
		\centering 
		\label{fig15_1}
	\end{center}
\end{figure}

Following this procedure, we do find a strong elongation of the droplets as Fig.~\ref{fig15_1} shown, where we compare the prediction of the full-equilibrium and this \textsl{restricted-equilibrium} model with the dynamical data of Metselaar \textsl{et al.} for tactoids of chitin in water. Shown is the aspect ratio of the droplets as a function of their volumes for a single electric field strength. For the largest droplets, the full relaxation takes more than the maximum of 1100 seconds, so the tactoids have not fully equilibrated yet, see Fig.~\ref{fig15_1} of \cite{Metselaar2017}.

\begin{figure}[h]
	\begin{center}
		\includegraphics[width=8cm]{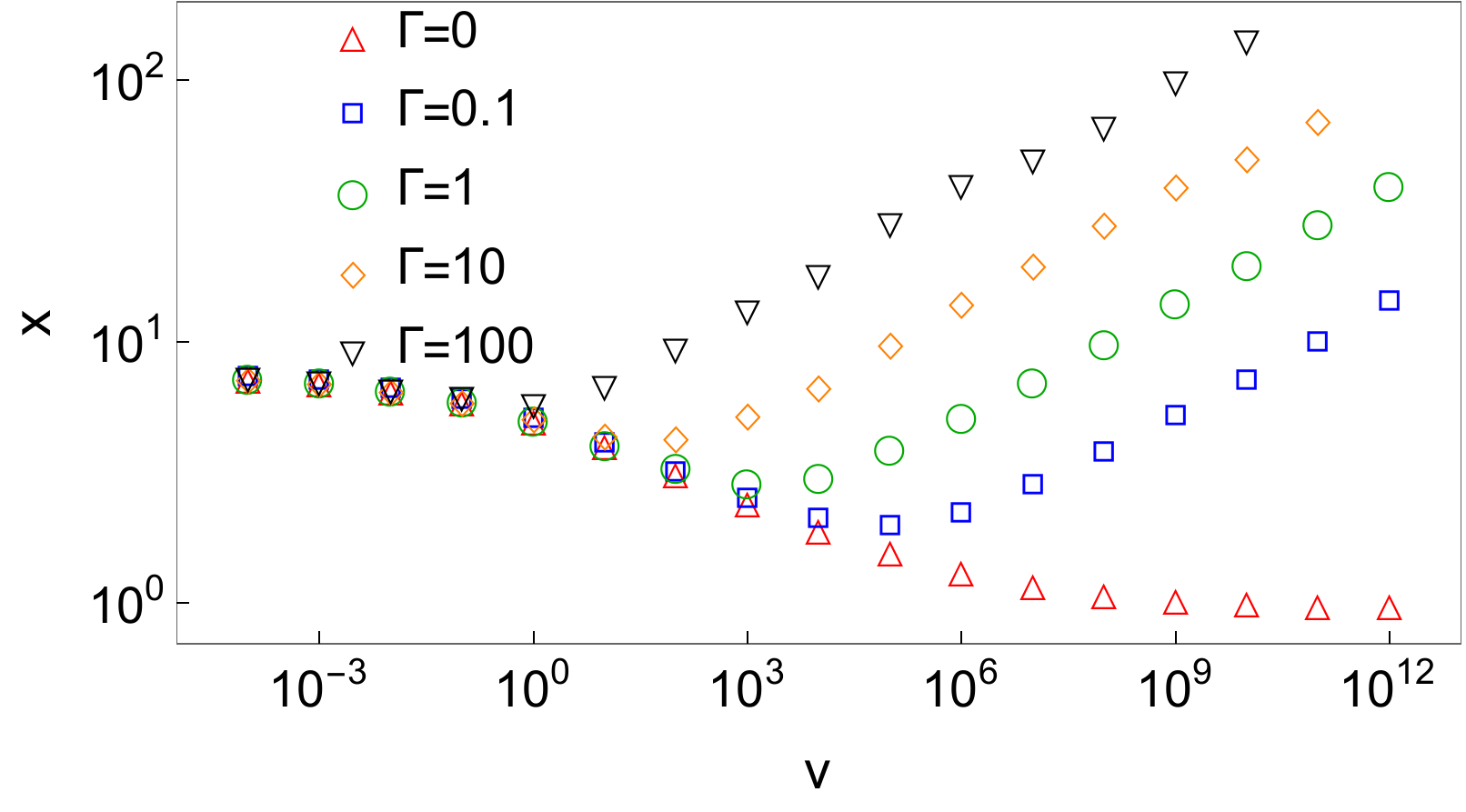}
		\caption{Aspect ratio $x$ of a tactoid as a function of the dimensionless volume $v$ for different electric field strengths $\Gamma$ according to the restricted equilibrium model. See the main text. Anchoring strength $\omega = 14$ and bend elastic constant $\kappa = 10$.}	
		\centering 
		\label{fig15}
	\end{center}
\end{figure}

\begin{figure}[h]
	\begin{center}
		\includegraphics[width=8cm]{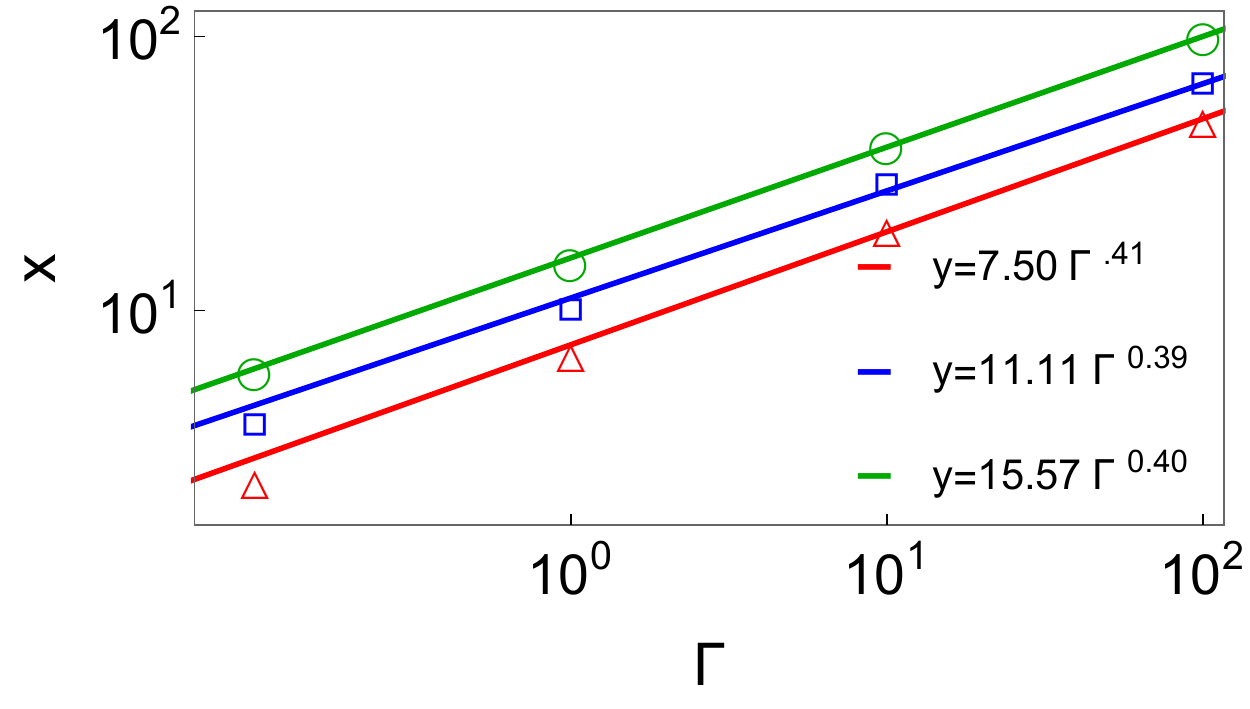}
		\caption{Aspect ratio $x$ of a tactoid as a function of the electric field $\Gamma$ according to the restricted equilibrium model. Anchoring strength $\omega =14$ and dimensionless bend constant $\kappa=10$. Different symbols show  different volumes: triangles $v=10^7$, squares $v=10^8$  and circles $v=10^9$. Indicated are also the scaling exponents, which are close to $0.40$ for the three tactoid volumes.}	
		\centering 
		\label{fig16}
	\end{center}
\end{figure}

It seems that within a restricted-equilibrium calculation, agreement with the experimental data is indeed rather good, even if they do not yet represent fully relaxed tactoids. The data confirm our expectation that the electric field only has an impact on the shape of the tactoids if they are sufficiently large. How large, depends on the strength of the electric field. This is shown in Fig.~\ref{fig15}, where we show predictions of our restricted-equilibrium model for the aspect ratio $x$ of nematic droplets as a function of the dimensionless volume $v$ for different dimensionless field strengths $\Gamma$. According to the scaling theory of section III, we should expect an $x\sim \Gamma^{3/7}v^{1/7}$ for a fully bipolar director field corresponding to sufficiently large droplets. The slopes of the various curves shown in Fig.~\ref{fig15} agree with this. Figure \ref{fig16} shows that the scaling with the electric field strength $\Gamma$ for different tactoid volumes $v$ also agrees with the scaling prediction of $3/7\approx 0.43$.

All of this of course begs the question why our full equilibrium model, in which the tactoids choose their optimal aspect ratio and director field in response to the external field, does not agree with the experimental observations. Above we have presumed that the bipolarness of the tactoids cannot respond swiftly to the switching on of an electric field, at least less swiftly than the aspect ratio can respond. In that case a restricted equilibrium picture applies, which would be valid for intermediate times. This implies that after an initial increase in aspect ratio, this aspect ratio should decrease again for (potentially) much later times. This has not yet been investigated but would be an interesting avenue of future experimental research. 

Whilst this may seem a somewhat far-fetched explanation to align theory and experiment, it does tie in with the observations of Jamali \textsl{et al.}, who collected data on hundreds of tactoids of carbon nanotubes in chlorosulfonic acid \cite{Jamali2015}. Even after $15$ days of equilibration, the scatter in the observed aspect ratio remains appreciable and cannot be explained by thermal fluctuations. Indeed, the experiments of Metselaar \textsl{et al.} also point at long relaxation times: the largest droplets do not seem completely equilibrated even after 7000 seconds. On the other hand, the Lattice Boltzmann simulations presented in the work of Metselaar \textsl{et al.} \cite{Metselaar2017}, which do mirror the large elongation of the tactoids in an external field, point at a relatively swift relaxation of the director field after the external field is switched on. 

In the simulations, the director field seems to keep the almost perfect planar alignment to the interface of the tactoid with the surrounding isotropic fluid, while in the bulk of the tactoid the director field seems to become homogeneous \cite{Metselaar_private}. 
This suggests a different kind of relaxation of the director field in response to the alignment field than the one we presumed in our work, which conserves the geometry of director field. This kind of director field is in our view surprising, as it involves a strong deformation with a small radius of curvature that is very costly in elastic free energy \cite{Siyu2018,Rudi2020}. This is why, generally, it is believed that wall defects in nematics spread out very quickly \cite{de1993}. (See, however, Tromp \textsl{et al.} \cite{tromp1996}.)  We should, in our view, not exclude the possibility that the simulations, which are coarse-grained and characterized by rather large interfacial widths even on the scale of the width of the droplets, allow for large deformations in the interfacial region. Because of this, we feel that additional and more comprehensive simulation studies would be useful to perform in order to settle this issue \cite{kuhnholdsimulating}. 

In fact, a simple scaling theory supports this view in the context of tactoids. Let us for simplicity take a spherical tactoid of radius $R$. A locally deformed director field that preserves perfect planar anchoring would give a free energy of the form $F \simeq \sigma R^2 + K R^2 \xi^{-1} + \gamma R^2 \xi$. Here, $K$ is some combination of the bend and splay elastic constants, $\xi \leq R$ is the width of the deformed director field that we equate to its radius of curvature, and $\gamma = \epsilon_a E^2$ is the Coulomb energy per unit volume. If we optimize $\xi$, we get $\xi = K_{11}^{1/2} \gamma^{-1/2} \leq R$ for $\gamma \geq K_{11} R^{-2}$. For $\gamma \leq K R^{-2}$, we have $\xi = R$.

Hence, we obtain $ F \simeq \sigma R^2 + KR + \gamma R^3$ for $\gamma \leq K R^{-2}$ and $ F \simeq \sigma R^2+\gamma^{1/2} + K^{1/2} R^2 $ for $\gamma \geq K R^{-2}$. For a smooth director field in the limit of large field strengths, we have $F \simeq \sigma R^2 + \sigma \omega R^2$ because the director field is then approximately uniform. This shows that for $\gamma \geq \omega \sigma /R$ the uniform director field has a lower free energy than the locally deformed one. Of course, we cannot exclude the possibility that for $K R^{-2}< \gamma < \omega \sigma/R$ a locally deformed director field wins out albeit that this might also be accompanied by an imperfect anchoring.

In conclusion, we should perhaps not exclude the possibility that the lattice Boltzmann simulations, which are coarse-grained and characterized by rather large interfacial widths even on the scale of the width of the droplets, allow for larger deformations in the interfacial region than a continuum theory would. Because of this, we feel that additional and more comprehensive simulation studies would be useful to perform in order to settle this issue \cite{kuhnholdsimulating}. 

Finally, we cannot exclude the possibility that the external field has a sizeable impact on both the interfacial tension, the anchoring and on the elastic constants, because they all depend on the degree of orientation order of the particles \cite{straley1973,odijk1986,van1999}.  Indeed, all of them depend on the degree of alignment of the particles, where we note that the isotropic phase becomes paranematic in the presence of an external field \cite{lee1987comment,khokhlov1982,szalai1998external}.  This implies that the interfacial tension between the nematic droplets and the host phase should decrease with increasing field strength. In fact, it should disappear altogether at some critical field strength. The study done in this paper shows that these issues can only be resolved with more detailed experimental investigation of the impact of external fields on the properties of isotropic and nematic phases of rod-like colloidal particles. 

This work was supported by the National Science Foundation through Grant No. DMR-1719550.

\section{Appendices}

\subsection{Bispherical coordinates}

\begin{figure}[!h]
	\begin{center}
		\includegraphics[width=5cm]{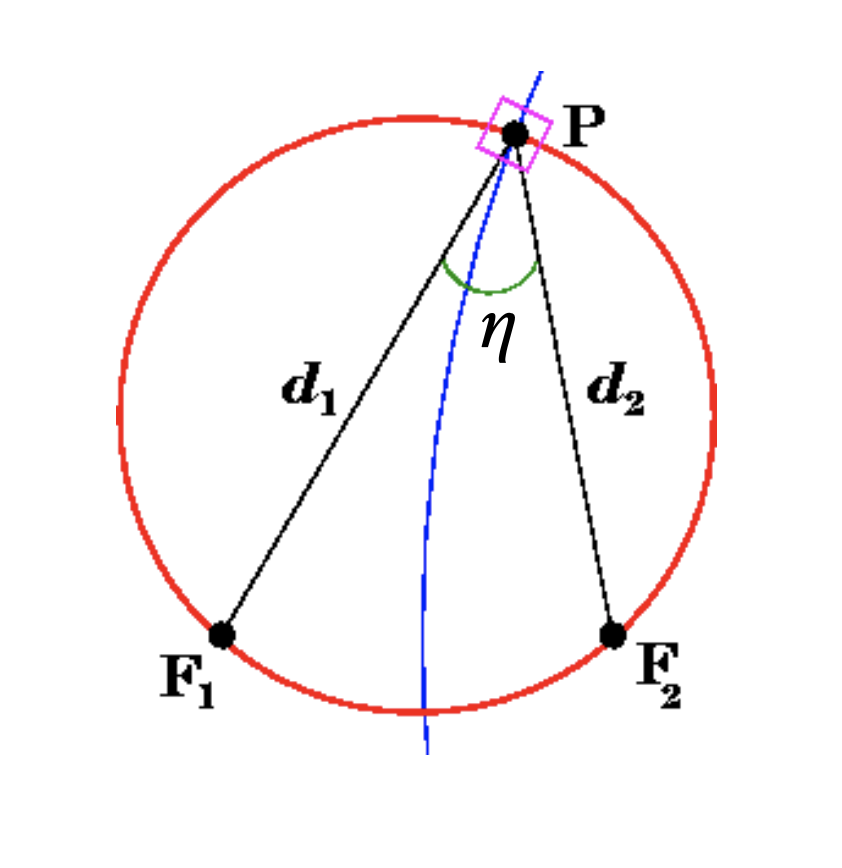}
		\caption{Parameters in bipolar coordinates: Each point $P$ on a curve can be described by  two parameters $\tau$ and $\eta$.  The angle $\eta$ at point $P$ sees the two foci while $\tau$ is the logarithm of the ratio of distances to two fixed (focal) points. The values of $\eta$ and $\tau$ are constant on the red and blue circles, respectively. The three-dimensional bispherical coordinate can be obtained by rotating the bipolar coordinate around the axis connecting the two focal points $F_1$ and $F-2$.  The third coordinate in the bispherical coordinate is denoted by the azimuthal angle $\phi$ not shown in the figure. Adopted from \cite{wiki}}	
		\centering 
		\label{bipolar_cor}
	\end{center}
\end{figure}

  In this section, we describe the bispherical coordinates used in the paper to define the structure of our tactoids \cite{williams1986,lucht2015bipolar,Prinsen2003,Kaznacheev2002}.  Bispherical coordinates are a three-dimensional orthogonal coordinate obtained from rotating the two-dimensional bipolar coordinate system about the axis that connects the two foci $F_{1}$ and  $F_{2}$, see Figs.~ 1-3 below.  We note that there is more than one way to define bipolar coordinates.  Within a bipolar coordinate system, every point $P$ on a curve can be described by two parameters that we denote $\tau$ and $\eta$. Here, $\tau= \log({d_1}/{d_2})$ with $d_1$ and $d_2$ the distance from each focal point $F_1$ and $F_2$ to that point on the curve, see Fig.~\ref{bipolar_cor}. As shown in the figure, $\eta$ corresponds to the angle that sees these two focal points.  
\\
The element of a surface in bispherical coordinate can be written as $dA= h_{\eta} h_{\tau} d \eta d \tau$, where
\begin{equation}
h_{\eta}=h_{\tau}= \dfrac{a}{\cosh \tau-\cos\eta }
\label{metrich}
\end{equation}
with $2a$ the distance between two focal points.  The rest of this section focuses on the derivation of the metric given in Eq.~\ref{metrich}.

\begin{figure}[!h]
	\begin{center}
		\includegraphics[width=6cm]{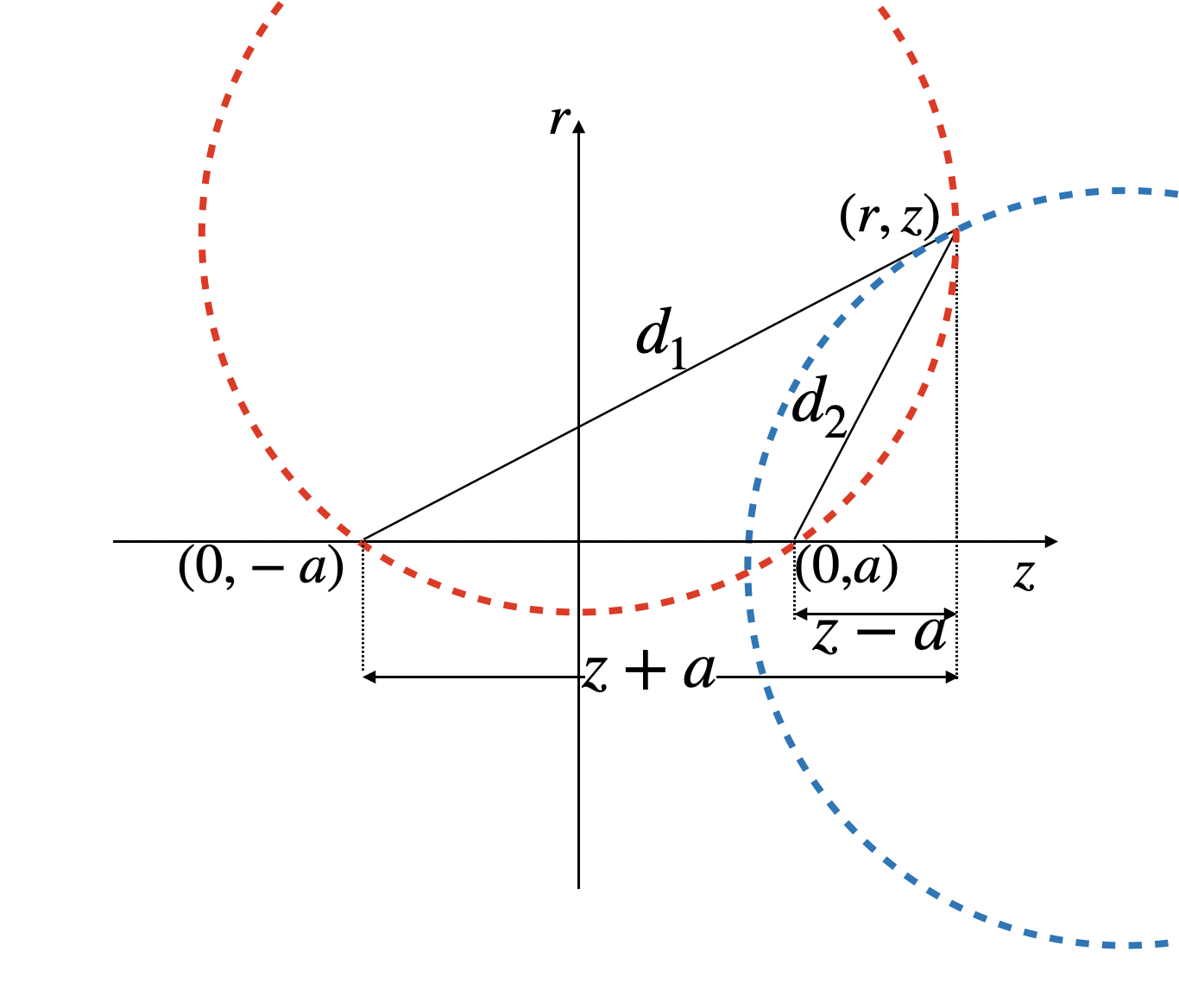}
		\caption{The relation between cylindrical coordinates and bispherical coordinates. For simplicity, we assume a point on the curve in the $\phi=0$ plane.}
		\centering 
		\label{fig:cylindrical}
	\end{center}
\end{figure} 

We first find the relation between the bispherical and cylindrical coordinates based on Fig.~\ref{fig:cylindrical}. For simplicity, we assume a point on the curve in the $\phi=0$ plane and write,
\begin{equation}
\begin{aligned}
& d_2^2=\left(z-a\right)^2+r^2, 
\\
& d_1^2=\left(z+a\right)^2+r^2. 
\end{aligned}
\label{eq34}
\end{equation}

To obtain $z$ and $r$ as a function of $\tau$ and $\eta$, we set $d_1 = d_2 e^\tau$ and find,

\begin{equation}
\begin{aligned}
& z=\dfrac{a \sinh \tau }{\cosh \tau -\cos \eta },
\\
& r = \dfrac{a\sin \eta }{\cosh \tau -\cos \eta }.
\label{eq36}
\end{aligned}
\end{equation}

\begin{figure}[!h]
	\begin{center}	\includegraphics[width=8cm]{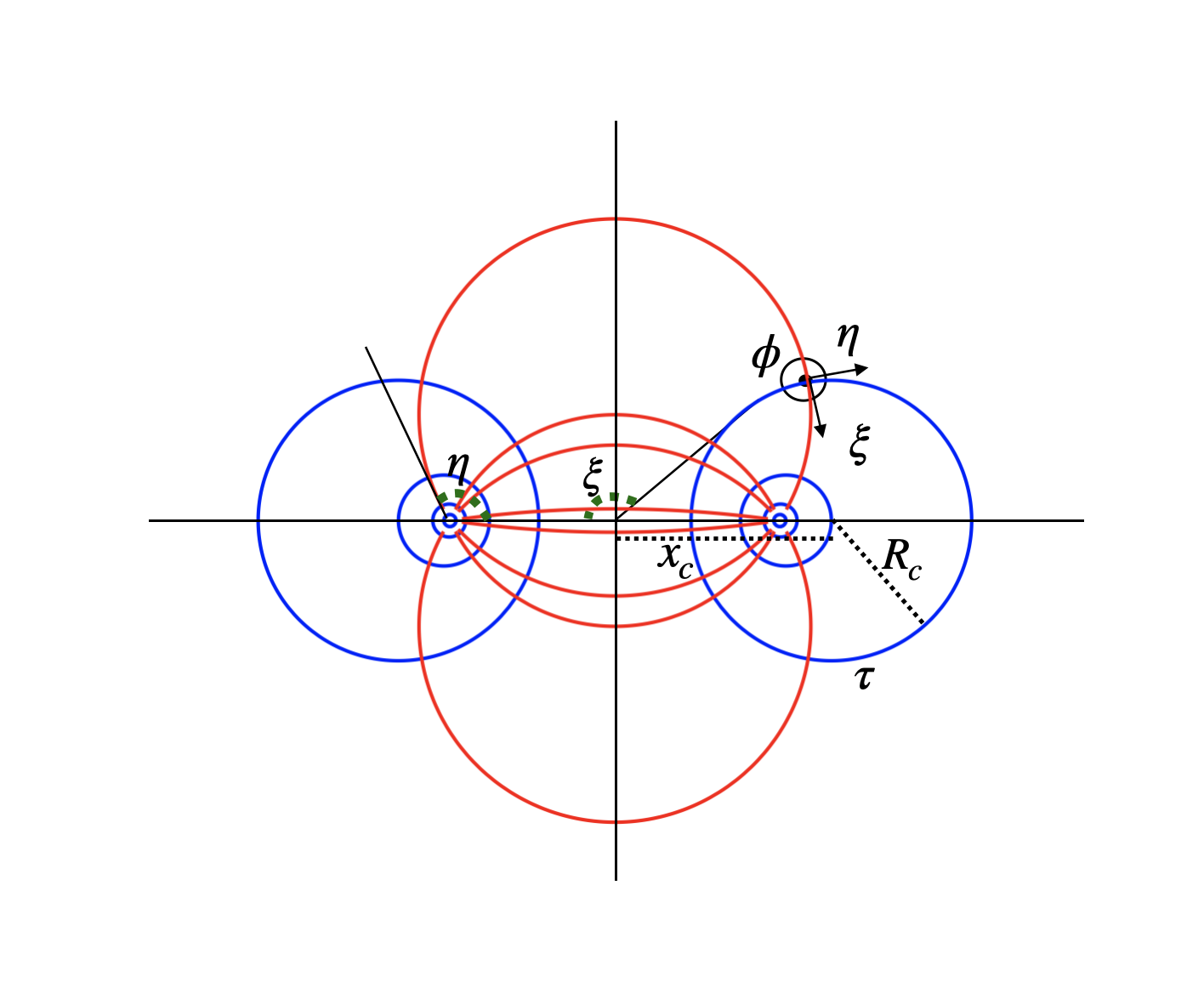}
		\caption{ Both $\xi$ and $\tau$ along with $\eta$ are constant values on blue circles and red arcs respectively. $x_c$ is the $x$-component of the right larger blue circle from the origin and the $R_c$ is the radius of the same circle.}	
		\centering 
	\label{fig:bisphericalxi}
	\end{center}
\end{figure}

Since the azimuthal angle $\phi$ is the same in both cylindrical and bispherical coordinates, we can easily obtain the relation between Cartesian and bispherical coordinates as follows,
\begin{equation}
\begin{cases}
x= \dfrac{a \sin \eta \cos \phi }{\cosh \tau -\cos \eta } \quad	\quad 0 < \eta < \pi
\\
y= \dfrac{a \sin \eta \sin \phi }{\cosh \tau -\cos \eta } \quad -\infty < \tau < \infty
\\
z= \dfrac{a \sinh \tau }{\cosh \tau -\cos \eta } \quad  \quad 0 < \phi < 2 \pi
\label{eq39}
\end{cases}
\end{equation}
We note that the value of $\tau$ is constant in blue circles shown in Fig~\ref{fig:bisphericalxi}. 
\\
 
It is straightforward to show that the center and radius of each circle are $x_c={a}/{\tanh \tau }$ and $R_c={a}/{\sinh \tau }$, respectively.

Instead of the parameter $\tau$, one can use the angle $\xi$ presented in Fig.~\ref{fig:bisphericalxi} with $\cot \xi ={\sqrt{x_c^2-R_c^2}}/{R_c} $, see Fig.~\ref{fig:bisphericalxi}. So, we can replace instead of $x_c$ and $R_c$ their functions.
Now, the Cartesian coordinate $x$, $y$ and $z$ can be defined as a function of bispherical coordinate $(\xi,\eta,\phi)$
\begin{equation}
\begin{cases}
x= \dfrac{a \sin \eta  \sin \xi  \cos \phi }{1+ \sin \xi \cos \eta } \quad	 0 < \eta < \pi
\\
y= \dfrac{a \sin \eta \sin \xi  \sin \phi }{1 + \sin \xi \cos \eta } \quad 0 < \xi < \pi
\\
z= \dfrac{a \cos \xi }{1+ \sin \xi  \cos \eta } \quad \ \ \   0 < \phi < 2 \pi
\label{eq43}
\end{cases}
\end{equation}
Since the bipoloar coordinate system is orthogonal, its metric tensor is diagonal.

Using Eq.~\ref{eq43}, we can find the diagonal terms $h_{\xi}^{2}=g_{\xi \xi}$ as follows,
\begin{equation}
h_{\xi}^{2}=g_{\xi \xi} = \sum_{k} \left(\dfrac{\partial X_k}{\partial \xi}\right)^2 =\left(\dfrac{\partial x}{\partial \xi}\right)^2 + \left(\dfrac{\partial y}{\partial \xi}\right)^2 +\left(\dfrac{\partial z}{\partial \xi}\right)^2.
\label{eq45}
\end{equation}
The other diagonal components can be calculated the same way. Finally, we find,
\begin{equation}
\begin{cases}
h_\xi = \dfrac{a}{1+\sin \xi \cos \eta },
\\
h_\eta =  \dfrac{a \sin \xi }{1+\sin \xi \cos \eta },
\\
h_\phi =  \dfrac{a \sin \xi  \sin \eta }{1+\sin \xi \cos \eta }.
\label{eq46}
\end{cases}
\end{equation}

\subsection{Some useful relations in the bispherical coordinate}
To obtain the free energy of a tactoid, we need to calculate
$\left(\div{{\bf n}}\right)$, 
$\left({\bf n}\cdot\left(\curl{\bf n}\right)\right)$
, 
$\left({\bf n}\times\left(\curl{\bf n}\right)\right)$
and 
$\left( \div{\left( {\bf n} \div{\bf n}+{\bf  n}\times\left(\curl{\bf n}\right)\right)} \right)$ with {\it {\bf n}} the director field.
Consider a general vector ${\bf A} = A_1 \hat{u}_1 + A_2 \hat{u}_2 +A_3 \hat{u}_3$.  Its divergence in the bispherical coordinate reads
\begin{equation}
\begin{aligned}
\div{{\bf A}} &= \dfrac{1}{h_1h_2h_3}(\dfrac{\partial}{\partial u_1}(h_2h_3A_1)+\dfrac{\partial}{\partial u_2}(h_1h_3A_2)
\\
&+\dfrac{\partial}{\partial u_3}(h_1h_2A_3))
\end{aligned}
\end{equation}
and the curl is,
\begin{equation}
\begin{aligned}
\curl{{\bf A}} = &  \dfrac{1}{h_2h_3}(\dfrac{\partial}{\partial u_2}(h_3A_3)-\dfrac{\partial}{\partial u_3}(h_2A_2))\hat{u}_1
\\
+ & \dfrac{1}{h_1h_3}(\dfrac{\partial}{\partial u_3}(h_1A_1)-\dfrac{\partial}{\partial u_1}(h_3A_3))\hat{u}_2 
\\
+& \dfrac{1}{h_1h_2}(\dfrac{\partial}{\partial u_1}(h_2A_2)-\dfrac{\partial}{\partial u_2}(h_1A_1))\hat{u}_3
\end{aligned}
\end{equation}
 
The divergence and curl of vector $\bf{n}= 1  \hat{\xi} + 0  \hat{\eta}  + 0 \hat{\phi }$ then become
\begin{equation}
\begin{aligned}
\div{\bf n}  & = \dfrac{1}{h_\xi h_\eta h_\phi} \dfrac{\partial}{\partial \xi} \left(h_\eta h_\phi\right)
\\
& = \dfrac{\left(1+\sin \xi \cos \eta \right)^3}{a^3\sin^2 \xi \sin \eta } (\dfrac{\partial}{\partial \xi} \dfrac{a^2 \sin^2 \xi  \sin \eta }{(1+\sin \xi \cos \eta )^2})
\\
& = \dfrac{2}{a} \cot  \xi 
\label{eq47}
\end{aligned}
\end{equation}
\begin{equation}
\begin{aligned}
\curl{\bf n} & = \dfrac{1}{h_\xi h_\phi} \left(\dfrac{\partial}{\partial \phi} h_\xi\right) \hat{\eta} - \dfrac{1}{h_\xi h_\eta} \left( \dfrac{\partial}{\partial \eta} h_\xi \right) \hat{\phi}
\\
& = \dfrac{\left(1+\sin \xi \cos \eta \right)^2}{a^2 \sin \xi } \left( \dfrac{\partial}{\partial \eta} \dfrac{a}{1+\sin \xi  \cos \eta }\right) \hat{\phi}
\\
& = - \dfrac{\sin \eta }{a} \hat{\phi}
\label{eq48}
\end{aligned}
\end{equation}
 
We can easily seen that $\bf{n}\cdot\curl{\bf{n}}=0$ and that
\begin{equation}
    {\bf n} \times \curl{\bf n} = \dfrac{1}{a} \sin\eta \left( \hat{\xi} \times \hat{\phi}\right) = \dfrac{1}{a} \sin \eta  \hat{\eta}.
\end{equation}

\subsection{Scaling theory}
To obtain the scaling behavior of the free energy of a tactoid, we set the volume of the droplet $V=r^2R$ and its surface area $S=rR$. As in the main text, the bipolarness is denoted by $y=\tilde{R}/R$ and the aspect ratio by $x=R/r$. To find the free energy, we need to calculate the volume integral of $\left(\div{\bf{n}}\right)^2$ and $\abs{\bf{n} \times \curl{\bf{n}}}^2$ and ${\left(\bf{E} \cdot \bf{n}\right)^2}$ and the surface integral of ${\left(\bf q \cdot n \right)^2}$, in terms of $x$, $y$ and $V$.

Considering that $\div{\bf n}$ is proportional to $R / \tilde{R}^2$, the splay term can then be written as

\begin{equation}
    \int \left(\div{{\bf n}}\right)^2 \dd V \propto \dfrac{R^2}{\tilde{R}^4} r^2R 
    \propto y^{-4}x^{-1}V^{1/3}x^{-1/3},
\end{equation}
where we have used the relation $r \propto V^{1/3}x^{-1/3}$ between $r$ and $V$ in the last term. 
One can also argue that $\abs{\curl{\bf n}}$ is proportional to $r/\tilde{R}^2$ because as the width of the droplet increase, the contribution to the bending energy increases too. Further, the bending energy decreases if the director-field becomes more homogeneous.  Thus the bending term can be written as

\begin{equation}
    \int \abs{\bf{n} \cross \left(\curl{{\bf n}} \right)}^2 \dd V \propto 
    \dfrac{R^4}{\tilde{R}^4} \dfrac{r^3}{R^3}r \propto y^{-4} x^{-3} V^{1/3}x^{-1/3}.
\end{equation}

Next we need to calculate the scaling behavior of the ${\bf q \cdot n}$ term associated with the anchoring strength contribution to the free energy. In Sect.~V below, we show that  

\begin{equation}
    \begin{aligned}
       {\bf q \cdot n} \sim \left( \dfrac{r}{R}-\dfrac{Rr}{\tilde{R}^2}\right).
    \end{aligned}
\end{equation}
Thus the contribution of the anchoring strength term to the surface energy scales as,
\begin{equation}
    \begin{aligned}
        \int \omega \left({\bf q \cdot n}\right)^2 \dd S &\propto \omega rR \left( \dfrac{r}{R}-\dfrac{Rr}{\tilde{R}^2}\right)^2 
        \\
        &\propto \omega rR\left(\dfrac{r}{R}\right)^2\left(1-\dfrac{R^2}{\tilde{R}^2}\right)
        \\
        &\propto \omega V^{2/3} x^{-5/3} \left(1-y^{-2}\right).
    \end{aligned}
\end{equation}
Finally, we need to calculate the interaction of director-field with the electric field as follows,

\begin{equation}
\begin{aligned}
    - \frac{1}{8\pi} \epsilon_a \int \left({\bf E \cdot n}\right)^2 \dd V &\propto - \epsilon_a E^2 \dfrac{R^2r^2}{\tilde{R}^4} r^2R
    \\
    &\propto - \epsilon_a E^2 y^{-4}x^{-2} V,
\end{aligned}
\end{equation}
where we have used the expression obtained in Sect. V below for ${\bf E \cdot n}$ in the limit of large $\tilde R$. 

\subsection{Free energy in bispherical coordinates}

In this section, we derive Eqs.~5-13 in the main text.  We parametrize the director field $\bf{n}\equiv \bf{t}/\vert \bf{t} \vert$ in the cylindrical coordinate with $\bf{t} = 2 \rho z \hat{\rho} - (\tilde{R}^2+\rho^2-z^2) \hat{z}$. Here, as in the main text, $\tilde{R}$ is the distance between the (virtual) boojums, $R$ is the length of droplet and $\hat{\rho}$ and $\hat{z}$ are the unit vector along the radial direction and the main axis of the tactoid in a cylindrical coordinate system, respectively. In a bispherical coordinate system, we have $\rho = R \sin \xi \sin \eta /Z$ and $z = R \cos \xi /Z$ with $Z = 1+\sin \xi \cos \eta$.
 
 We first calculate the divergence and curl of ${\bf{n}}$ in the cylindrical coordinate and then transfer them to the bispherical one in order to obtain the free energy of a droplet illustrated in Fig.~2. In the cylindrical coordinate, we have 
 \begin{equation}
     \div{\bf{n}}=\dfrac{1}{\rho}\dfrac{\partial \rho n_\rho}{\partial \rho}+\dfrac{1}{\rho}\dfrac{\partial n_\varphi}{\partial \varphi}+\dfrac{\partial n_z}{\partial z}
 \end{equation}
 with
 \begin{equation}
 \begin{aligned}
    \bf{n} =& \dfrac{2\rho z}{\sqrt{4\rho^2 z^2+\left(\tilde{R}^2+\rho^2-z^2\right)^2}}\hat{\rho}
    \\
    &-\dfrac{\left(\tilde{R}^2+\rho^2-z^2\right)}{\sqrt{4\rho^2 z^2+\left(\tilde{R}^2+\rho^2-z^2\right)^2}}\hat{z}
\end{aligned}
 \end{equation}
 After some tedious calculation, we find
 \begin{equation}
     \begin{aligned}
          \div{\bf{n}}=&
         \dfrac{4z}{\left(4\rho^2 z^2+\left(\tilde{R}^2+\rho^2-z^2\right)^2\right)^{1/2}},
     \end{aligned}
 \end{equation}
 and 
 \begin{equation}
     \begin{aligned}
         &\left(\div{\bf{n}}\right)^2=\dfrac{16z^2}{4\rho^2 z^2+\left(\tilde{R}^2+\rho^2-z^2\right)^2}
         \\
         &= 
         \dfrac{4R^{-2}\cos^2\xi}{Z^2\left(\sin^2\xi\sin^2\eta\cos^2\xi/Z^4+ Y_{11} \right)}
         \\
         & Y_{11} = \left(\tilde{R}^2/2R^2+\sin^2\xi\sin^2\eta/2Z^2-\cos^2\xi/2Z^2\right)^2 .
     \end{aligned}\label{eq:divn2}
 \end{equation}
We set the denominator of the above equation equal to $N$, and simplify it to find

\begin{equation}
     \begin{aligned}
         N=&\sin^2\xi\sin^2\eta\tilde{R}^2/R^2
         \\
         &+\left(Z/2\left(\tilde{R}^2/2R^2-1\right)+\sin\xi\cos\eta\right)^2
     \end{aligned}
 \end{equation}
 It is straightforward to show that
 \begin{equation}
     N R^4=\abs{t}^2 Z^2,
     \label{eq:N}
 \end{equation}
 which we will use later.
 The curl in the cylindrical coordinate can be calculated as follows,
 \begin{equation}
     \begin{aligned}
          \curl{\bf{n}} = & \left(\frac{1}{\rho}\frac{\partial n_z}{\partial \varphi} - \frac{\partial n_\varphi}{\partial z}\right)\hat{\rho} + \left(\frac{\partial n_\rho}{\partial z} - \frac{\partial n_z}{\partial \rho}\right)\hat{\varphi} 
          \\ &+\frac{1}{\rho}\left(\frac{\partial}{\partial \rho}\left(\rho n_\varphi\right) - \frac{\partial n_\rho}{\partial \varphi}\right) \hat{z}, \\[8px]
     \end{aligned}
 \end{equation}
 
which becomes,
\begin{equation}
     \begin{aligned}
        \curl{\bf{n}}&=\dfrac{2\rho\left(\tilde{R}^2+\rho^2-z^2\right)^2+8\rho^3z^2}{\left(4\rho^2 z^2+\left(\tilde{R}^2+\rho^2-z^2\right)^2\right)^{3/2}}\hat{\varphi}
        \\
        &= \dfrac{2\rho}{\left(4\rho^2 z^2+\left(\tilde{R}^2+\rho^2-z^2\right)^2\right)^{1/2}}\hat{\varphi}.
     \end{aligned}
 \end{equation}
 Using the above equation, we can calculate the following term
\begin{equation}
    \begin{aligned}
        \abs{\bf{n}\times\left(\curl{\bf{n}}\right)}^2&=\dfrac{4R^2\sin^2\xi\sin^2\eta Z^{-2}}{\abs{t}^2}
        \\
        &=\dfrac{4R^2\sin^2\xi\sin^2\eta Z^{-2}}{NR^4Z^{-2}}
        \\
        &=\dfrac{4R^{-2}\sin^2\xi\sin^2\eta }{N},
        \label{eq:ncurln2}
    \end{aligned}
\end{equation}
which we will use later when we calculate the bending energy.

To calculate the contribution of anchoring strength to the free energy, we need to calculate the term $\left({\bf q}\cdot {\bf n}\right)^2$ with ${\bf q}$ the normal vector to the surface. Setting the unit tangent vector to the surface with ${\bf t'}$, we then find $\left({\bf t'}\times{\bf n}\right)^2=\left({\bf q}\cdot {\bf n}\right)^2$. We emphasize that ${\bf t'}$ is the tangent vector to the surface, not the tangent vector to the field lines that we denoted ${\bf t}$ above. The unit tangent vector ${\bf t'}$ can be written as ${\bf t'}\equiv {\bf t''}/\vert {\bf t''} \vert$ with ${\bf t''} = 2 \rho z \hat{\rho} - (R^2+\rho^2-z^2) \hat{z}$. Therefore, 
 \begin{equation}
     \begin{aligned}
         \abs{\bf{q}\cdot \bf{n}} = & \dfrac{\abs{\bf{t''}\times\bf{t}}}{\abs{\bf{t''}}\abs{\bf{t}}}
         \\
         = &
         \dfrac{\abs{\left(2 \rho z \hat{\rho} - \left(R^2+\rho^2-z^2\right) \hat{z}\right)}}{\sqrt{4 \rho^2 z^2 + \left(R^2+\rho^2-z^2\right)^2}}
         \\
         & \dfrac{\left(2 \rho z \hat{\rho} - \left(\tilde{R}^2+\rho^2-z^2\right) \hat{z}\right)}{\sqrt{4 \rho^2 z^2 - \left(\tilde{R}^2+\rho^2-z^2\right)^2}},
     \end{aligned}
 \end{equation}
 or, in the bispherical coordinate, we can finally write
\begin{equation}
    \begin{aligned}
        \abs{\bf{q}\cdot\bf{n}}^2 = \dfrac{4\sin^2\eta \cos^2\xi \left(\tilde{R}^2/R^2-1\right)^2 }{4N}.
        \label{eq:qn2}
    \end{aligned}
\end{equation}

 Last but not least, we need to calculate $\bf{E}\cdot\bf{n}$ to obtain the effect of the interaction of the external field $\bf{E}$ with the tactoid.  Considering that $\bf{n}={t}/\abs{\bf{t}}$ and assuming that the electric field is in the $z$ direction, we find
 
 \begin{equation}
     \begin{aligned}
         \dfrac{\left(\bf{E}\cdot\bf{t}\right)^2}{\abs{t}^2} =& \dfrac{E^2 \left(\tilde{R}^2+\rho^2-z^2\right)^2}{\abs{t}^2}
         \\
         =& \dfrac{E^2R^4Z^{-4}\left(Z^2\tilde{R}^2/R^2+\sin^2\xi\sin^2\eta-\cos^2\xi\right)^2}{NR^4Z^{-2}}
         \\
         =& \dfrac{E^2\left(Z^2\tilde{R}^2/R^2+\sin^2\xi\sin^2\eta-\cos^2\xi\right)^2}{N\left(1+\sin\xi\cos\eta\right)^{2}}.
         \label{eq:En2}
     \end{aligned}
 \end{equation}

To obtain the free energy given in Eq.~5 in the main text, we insert into Eq.~2 of the paper the expressions obtained above for $\left(\div{\bf{n}}\right)^2$ (Eq.~\ref{eq:divn2}), $\abs{\bf{n}\times\left(\curl{\bf{n}}\right)}^2$ (Eq.~\ref{eq:ncurln2}), $\abs{\bf{q}\cdot\bf{n}}^2$ (Eq.~\ref{eq:qn2}), and $\bf{E}\cdot\bf{n}$ (Eq.~\ref{eq:En2}).   
Note that the volume of a droplet in the bispherical coordinate can be written as,
\begin{equation}
    V\left(\alpha,R \right) = \int_0^{2\pi} \int_0^{\alpha} \int_0^{\pi}h_\xi h_\eta h_\phi \ \dd \xi \ \dd \eta \ \dd \phi = R^3 \phi_v\left(\alpha\right),
\end{equation}
where $\alpha$ is the angle shown in Fig.~2 in the main text, $\phi_\mathrm{v}(\alpha)$ is
\begin{equation}
\phi_\mathrm{v}(\alpha) = \frac{7\pi}{3}+ \frac{\pi}{2}\left( \frac{1 - 4\alpha \cot \alpha + 3\cos 2\alpha}{\sin^2 \alpha}\right),
\label{eq:phiv}
\end{equation} 
and the surface can be expressed as
\begin{equation}
    S\left(\alpha,R \right) = \int_0^{2\pi} \int_0^{\pi}h_\xi h_\phi \ \dd \xi \ \dd \phi = R^2, \phi_\sigma\left(\alpha\right)
\end{equation}
with
\begin{equation}
\begin{aligned}
\phi_\sigma(\alpha) = 4\pi \left(  \frac{1-\alpha \cot\alpha}{\sin \alpha}\right).
\label{eq:phis}
\end{aligned}
\end{equation}
Using Eq.~\ref{eq:qn2}, the term in the free energy originating from the anchoring of the director field to the interface is
\begin{equation}
\begin{aligned}
\phi_{\omega}(\alpha,y) =&\dfrac{1}{R^2} \int_0^{2\pi}\int_0^\pi \left(\bf{q}\cdot \bf{n} \right)^2 h_\xi h_\phi \ \dd \xi \ \dd \phi
\\
=& \frac{\pi}{2} (y^2-1)^2 \sin^3 \alpha
\\
&\int_{0}^{\pi} \mathrm{d} \xi \left[  \dfrac{\sin \xi  \cos^2 \xi   }{N(y,\xi,\alpha)\left(1+\sin \xi \cos \alpha \right)^2} \right],
\label{eq:phiomega}
\end{aligned}
\end{equation}
which is equal to Eq.~8 in the main text and cannot be solved analytically.

Note that we have set $\eta = \alpha$ in the expressions given in Eqs.~(8) and (9) in the paper.

Using Eq.~\ref{eq:divn2}, we can also calculate the contribution of the splay deformation to the Frank elastic energy as follows,  
\begin{equation}
\begin{aligned}
\phi_{11}(\alpha,y) =& \dfrac{1}{R} \int_0^{2\pi}\int_0^\pi\int_0^\alpha \left(\div{\bf{n}}\right)^2 h_\xi h_\eta h_\phi \ \dd \xi \ \dd \eta \ \dd \phi
\\
=& 8\pi  \int_{0}^{\pi} \mathrm{d} \xi \int_{0}^{\alpha} \mathrm{d} \eta  \frac{\sin^2 \xi  \cos^2 \xi  \sin \eta}{N(y,\xi,\eta)\left(1+\sin \xi \cos \eta \right)^3},
\label{eq:phi11}
\end{aligned}
\end{equation}
which cannot be solved analytically either.

Using Eq.~\ref{eq:ncurln2}, the contribution of the bend elastic deformation reads
\begin{equation}
\begin{aligned}
 \phi_{33}(\alpha,y) =& 
\dfrac{1}{R} \int_0^{2\pi}\int_0^\pi\int_0^\alpha \abs{\bf{n} \times \curl{\bf{n}}}^2 h_\xi h_\eta h_\phi \ \dd \xi \ \dd \eta \ \dd \phi
\\
=& 8\pi \int_{0}^{\pi} \mathrm{d} \xi \int_{0}^{\alpha} \mathrm{d} \eta   \frac{\sin^4 \xi  \sin^3 \eta }{N(y,\xi,\eta)\left(1+\sin \xi \cos \eta \right)^3}.
\label{eq:phi33}
\end{aligned}
\end{equation}

Finally, the free energy of the interaction of the nematic drop with an electric field yields an even more daunting integral,
\begin{equation}
\begin{aligned}
\phi_{\mathrm{C}}(\alpha,y)   &=\dfrac{1}{R^3}  \int_0^{2\pi}\int_0^\pi\int_0^\alpha \left(\bf{E}\cdot\bf{n}\right)^2 h_\xi h_\eta h_\phi \ \dd \xi \ \dd \eta \ \dd \phi
\\
    &= 8\pi \int_{0}^{\pi} \mathrm{d} \xi \int_{0}^{\alpha} \mathrm{d} \eta  \frac{\sin^2 \xi  \sin \eta }{(1+\sin \xi \cos \eta )^3}
    \\
 &\times \dfrac{E_z^2\left(Z^2\tilde{R}^2/R^2+\sin^2\xi\sin^2\eta-\cos^2\xi\right)^2}{N\left(1+\sin\xi\cos\eta\right)^{2}}.
 \label{eq:phic}
\end{aligned}
\end{equation}
\subsection{Small opening angle}
In this section we calculate the free energy in the limit of the small opening angle. We first expand $\phi_v\left(\alpha\right)$ and $\phi_\sigma\left(\alpha\right)$ given in Eqs.~\ref{eq:phiv} and \ref{eq:phis} in the limit of small $\alpha$,
\begin{equation}
\begin{aligned}
    \phi_v\left(\alpha\right)&=\dfrac{7\pi}{3}+\dfrac{\pi}{2\sin^2\alpha}\left(1-4\alpha\cot{\alpha}+3\cos{\left(2\alpha\right)}\right)
    \\
    &\sim \dfrac{4\pi\alpha^2}{15}
\end{aligned}
\end{equation}
and
\begin{equation}
    \phi_\sigma\left(\alpha\right)=4\pi\dfrac{1-\alpha\cot{\alpha}}{\sin{\alpha}}\sim\dfrac{4\pi \alpha}{3},
\end{equation}
and next calculate different contributions to the free energy. Using Eq.~\ref{eq:phiomega}, the term associated with the anchoring strength in the limit of small $\alpha$ becomes,
\begin{equation}
\begin{aligned}
\phi_{\omega}(\alpha,y) = & \frac{\pi}{2} \left(y^2-1\right)^2 \sin^3 \alpha  
\\
&\int_{0}^{\pi} \mathrm{d} \xi \left[  \dfrac{\sin \xi  \cos^2 \xi   }{N(y,\xi,\alpha)\left(1+\sin \xi \cos \alpha \right)^2} \right]
\\
 \sim&\left(y^2-1\right)^2\sin^3\alpha\int_0^\pi\dd \xi \dfrac{\sin{\xi}\cos^2{\xi}}{\left(1+\sin{\xi}\right)^4y^4}
\\
 \sim & \left(y^2-1\right)^2y^{-4}\alpha^3 
\\
 \sim & \left(1-y^{-2}\right)^2\alpha^3.
\label{eq:phiomegasmall}
\end{aligned}
\end{equation}
We employ Eq.~\ref{eq:phiomegasmall} to calculate the term associated with the anchoring strength given in Eq.~5 of the paper in the limit of small $\alpha$,
\begin{equation}
\begin{aligned}
\dfrac{\phi_\omega\left(\alpha,y\right)}{\phi_v^{2/3}\left(\alpha\right)}& \sim \left(1-y^{-2}\right)^2\alpha^3 \alpha^{-4/3} 
\\
& \sim \left(1-y^{-2}\right)^2\alpha^{5/3} \sim \left(1-y^{-2}\right)^2 x^{-5/3}.
\end{aligned}
\end{equation}
Using Eq.~\ref{eq:phi11} in the limit of small $\alpha$, we find,
\begin{equation}
\begin{aligned}
\phi_{11}(\alpha,y) & = 8\pi  \int_{0}^{\pi} \mathrm{d} \xi \int_{0}^{\alpha} \mathrm{d} \eta \frac{\sin^2 \xi  \cos^2 \xi  \sin \eta}{N(y,\xi,\eta)\left(1+\sin \xi \cos \eta \right)^3}
\\
& \sim y^{-4}\alpha^{2} \int_0^\pi\dfrac{\sin^2\xi\cos^2\xi}{y^{4}\left(1+\sin\xi\right)^5}
\\
& \sim  y^{-4}\alpha^{2}.
\label{eq:phi11small}
\end{aligned}
\end{equation}
We employ Eq.~\ref{eq:phi11small} to calculate the contribution of the splay term to Eq.~5 in the paper,
\begin{equation}
\begin{aligned}
\dfrac{\phi_{11}\left(\alpha,y\right)}{\phi_v^{1/3}\left(\alpha\right)} & \sim y^{-4} \alpha^{4/3} \sim y^{-4} x^{-4/3}.
\end{aligned}
\end{equation}
Next, to obtain the contribution of the bending term, we calculate Eq.~\ref{eq:phi33} in the limit of small $\alpha$:
\begin{equation}
\begin{aligned}
\phi_{33}(\alpha,y) & = 8\pi \int_{0}^{\pi} \mathrm{d} \xi \int_{0}^{\alpha} \mathrm{d} \eta   \frac{\sin^4 \xi  \sin^3 \eta }{N(y,\xi,\eta)\left(1+\sin \xi \cos \eta \right)^3}
\\
& \sim  y^{-4}\alpha^4\int_0^\pi \dfrac{\sin^4\xi}{\left(1+\sin\xi\right)^5}
\\
& \sim y^{-4}\alpha^{4},
\end{aligned}
\label{eq:phi33small}
\end{equation}
and use it to find the contribution of bending energy to Eq.~5 of the paper as follows:
\begin{equation}
\begin{aligned}
\dfrac{\phi_{33}\left(\alpha,y\right)}{\phi_v^{1/3}\left(\alpha\right)}& \sim y^{-4}\alpha^{10/3} \sim y^{-4}x^{-10/3}.
\end{aligned}
\end{equation}
Finally, the interaction with the electric field $\int \left(\bf{E}\cdot\bf{n}\right)^2\dd V$ in the limit of small $\alpha$ becomes,
\begin{equation}
    \begin{aligned}
       -\int &\left(\bf{E}\cdot\bf{n}\right)^2\dd V
       =-\int \dfrac{E_z^2 \left(\tilde{R}^2+\rho^2-z^2\right)^2}{4\rho^2z^2+\left(\tilde{R}^2+\rho^2-z^2\right)^2} \dd V 
       \\
       =&
       -\int E_z^2 \left(1+\dfrac{4\rho^2z^2}{\left(\tilde{R}^2+\rho^2-z^2\right)^2}\right)^{-1}\dd V
       \\
      \sim &-\int E_z^2\left(1-\dfrac{4\sin^2\xi\cos^2\xi\sin^2\eta}{\left(Z^2\tilde{R}^2/R^2+\sin^2\xi\sin^2\eta-\cos^2\xi\right)^2}\right)\dd V
      \\
      \sim&
      -E_z^2R^3\alpha^4\int_0^\pi \dfrac{4\sin^2\xi\cos^2\xi}{\tilde{R}^4/R^4}\dfrac{\sin^2\xi}{\left(1+\sin\xi\right)^3} \dd \xi 
      \\
      \sim& - V E_z^2 x^{-2} y^{-4},
    \end{aligned}
\end{equation}
where the dropped the integral as it is just a number. All equations obtained in this section in the limit of the small opening angle are in agreement with the scaling theory presented in the main text.

\bibliography{arxivmain}
\end{document}